\documentclass[12pt,prd,onecolumn,showpacs,amsmath,amssymb,aps,floats,floatfix]{revtex4-1}

\usepackage[colorlinks=true,urlcolor=blue,anchorcolor=blue,citecolor=blue,filecolor=blue,linkcolor=blue,menucolor=blue,linktocpage=true]{hyperref} 

\usepackage[inline]{enumitem}
\usepackage[multidot]{grffile}  
\usepackage{dcolumn}
\usepackage{bm}
\usepackage{amsmath}
\usepackage{amsfonts}
\usepackage{amssymb}
\usepackage{color}
\usepackage[table]{xcolor}
\usepackage{float}
\usepackage{latexsym}
\usepackage{slashed} 
\usepackage{pstricks}
\usepackage{indentfirst}
\usepackage{mathrsfs}
\usepackage{multirow}
\usepackage{epsfig,psfrag}
\usepackage{subfigure}
\usepackage{mathtools}
\usepackage{setspace} 
\usepackage[utf8]{inputenc} 
\usepackage[scientific-notation=true]{siunitx} 

\graphicspath{{fig/}}

\def\pT{p_\mathrm{T}} 
\def\missET{\slashed E_\mathrm{T}} 
\def\misspT{\slashed{\mathbf{p}}_\mathrm{T}} 
\def\mT{m_\mathrm{T}}
\def\mT2{m_\mathrm{T2}}

\newcommand{\SUtwo}{\mathrm{SU}(2)}
\newcommand{\SUtwoL}{\mathrm{SU}(2)_\mathrm{L}}
\newcommand{\UoneY}{\mathrm{U}(1)_\mathrm{Y}}

\newcommand{\SUtwoR}{\mathrm{SU}(2)_\mathrm{R}}

\allowdisplaybreaks 

\begin{document}

\title{Inert sextuplet scalar dark matter at the LHC and future colliders}
\author{Dan-Yang Liu}
\author{Chengfeng Cai}
\author{Zhao-Huan Yu}\email[Corresponding author. ]{yuzhaoh5@mail.sysu.edu.cn}
\author{Yu-Pan Zeng}
\author{Hong-Hao Zhang}\email[Corresponding author. ]{zhh98@mail.sysu.edu.cn}
\affiliation{School of Physics, Sun Yat-Sen University, Guangzhou 510275, China}

\begin{abstract}

We study a dark matter model constructed by extending the standard model with an inert $\mathrm{SU}(2)_\mathrm{L}$ sextuplet scalar of hypercharge 1/2. The sextuplet components are split by the quartic couplings between the sextuplet and the Higgs doublet after electroweak symmetry breaking, resulting in a dark sector with one triply charged, two doubly charged, two singly charged, and two neutral scalars. The lighter neutral scalar boson acts as a dark matter particle. We investigate the constraints on this model from the $\text{monojet} + \missET$ and $\text{soft-dilepton} + \text{jets} + \missET$ searches at the 13~TeV Large Hadron Collider, as well as from the current electroweak precision test. Furthermore, we estimate the projected sensitivities of a 100~TeV $pp$ collider and of a future $e^+e^-$ collider, and find that such future projects could probe TeV mass scales.
Nonetheless, such mass scales only correspond to a subdominant component of the observed relic abundance if the dark matter particles solely originate from thermal production.
\end{abstract}

\maketitle
\tableofcontents
\clearpage

\section{Introduction}

The standard model (SM) of particle physics is a self-consistent $\mathrm{SU(3)_{C}\times SU(2)_{L}\times U(1)_{Y}}$ gauge theory describing the properties and interactions of three generations of fundamental fermions~\cite{Glashow:1961tr,Weinberg:1967tq,Salam:1968rm}.
It has been well tested by a variety of experiments, and all the fundamental particles it predicts have been found.
However, the SM cannot explain the existence of cold dark matter (DM) in the Universe, which is strongly suggested by astrophysical and cosmological experiments~\cite{Bertone:2004pz,Feng:2010gw,Young:2016ala}.
In order to account for particle dark matter, the SM must be extended.

Many popular extensions involve weakly interacting massive particles (WIMPs) as DM candidates, since thermal production of WIMPs in the early Universe is typically consistent with the observation of DM relic abundance.
From the viewpoint of model building, WIMP models can be directly established by introducing extra colorless $\mathrm{SU}(2)_\mathrm{L}$ multiplets as the dark sector~\cite{Mahbubani:2005pt,Cirelli:2005uq,Barbieri:2006dq,Gustafsson:2007pc,Cirelli:2007xd,Cao:2007rm,FileviezPerez:2008bj,Cirelli:2009uv,Hambye:2009pw,Araki:2011hm,Cohen:2011ec,Cai:2012kt,JosseMichaux:2012wj,Earl:2013jsa,AbdusSalam:2013eya,Fischer:2013hwa,Earl:2013fpa,Dedes:2014hga,Ayazi:2014tha,Harigaya:2015yaa,Ostdiek:2015aga,Cirelli:2015bda,Garcia-Cely:2015dda,Cai:2015kpa,Tait:2016qbg,Banerjee:2016hsk,Khan:2016sxm,Logan:2016ivc,Lu:2016dbc,Cai:2016sjz,Chowdhury:2016mtl,Cai:2017wdu,Liu:2017gfg,Xiang:2017yfs,Wang:2017sxx,
Cai:2017fmr,Lopez-Honorez:2017ora,Cai:2018nob,DuttaBanik:2018emv,Gu:2018kmv,Betancur:2018xtj,Kadota:2018lrt,Wang:2018lhk,Filimonova:2018qdc,Chao:2018xwz,Abe:2019wku,Cai:2019mtu,Cheng:2019qbd,Zeng:2019tlw,Bell:2020gug,Chiang:2020rcv,Jangid:2020qgo,Konar:2020vuu}.
The lightest mass eigenstate from the electrically neutral components of the multiplets could be a viable DM candidate, which can be either a fermion or scalar boson.
In order to ensure the stability of the DM candidate, a $Z_2$ symmetry is commonly required.

In this way, scalar DM models can be constructed by extending the scalar sector with an inert scalar multiplet that does not develop a vacuum expectation value (VEV) because of an unbroken $Z_2$ symmetry.
Such extensions with doublet~\cite{Deshpande:1977rw,Barbieri:2006dq,Gustafsson:2007pc,Cao:2007rm}, triplet~\cite{Cirelli:2007xd,FileviezPerez:2008bj,Hambye:2009pw,Araki:2011hm,JosseMichaux:2012wj,AbdusSalam:2013eya,Ayazi:2014tha,Khan:2016sxm,Bell:2020gug,Chiang:2020rcv,Jangid:2020qgo}, quadruplet~\cite{AbdusSalam:2013eya,Chowdhury:2016mtl,Cai:2017wdu,Zeng:2019tlw}, quintuplet~\cite{Cirelli:2007xd,Hambye:2009pw,Earl:2013jsa,Cai:2017fmr,Chao:2018xwz}, sextuplet~\cite{Earl:2013jsa,Earl:2013fpa,Logan:2016ivc}, septuplet~\cite{Cirelli:2007xd,Hambye:2009pw,Cai:2012kt,Earl:2013jsa,Garcia-Cely:2015dda,Cai:2015kpa,Chao:2018xwz}, and octuplet~\cite{Earl:2013jsa,Earl:2013fpa,Logan:2016ivc} scalars have been discussed in the literature.

In this work, we focus on the case of an inert sextuplet scalar with hypercharge $Y=1/2$.
A previous study~\cite{Earl:2013fpa} on this case investigated the phenomenological constraints from unitarity, electroweak oblique parameters, loop-induced Higgs decays into $\gamma\gamma$ and $Z\gamma$, DM relic abundance, and DM direct detection.
A follow-up work~\cite{Logan:2016ivc} analyzed the impact of coannihilation and Sommerfeld enhancement effects on the relic abundance calculation, as well as the scale where a Landau pole appears.
Here we extend these studies to the searches for the  sextuplet scalar at the Large Hadron Collider (LHC) and future colliders, in order to establish a more comprehensive phenomenological picture for such a model.

After electroweak symmetry breaking, the mass eigenstates from the sextuplet scalar involve one triply charged scalar, two doubly charged scalars, two singly charged scalars, and two neutral scalars.
Thus, the model predicts a series of fundamental scalar bosons, whose number and electric charges are determined by the gauge representation of the sextuplet.
This is a distinct, exotic prediction for future experimental searches.
The lighter neutral scalar boson acts as a DM candidate.
There are particular flat directions among the scalar couplings resulting in vanishing DM-nucleon scattering cross section at tree level.
This provides a possible explanation for the null search results in recent direct detection experiments.
At a high energy $pp$ collider, the scalar bosons could be directly produced in pairs via electroweak interactions.
Their decay chains would end at the DM candidate, which escapes detection and typically leaves a large missing transverse energy ($\missET$).
Potential searching channels in $pp$ collisions include the $\text{monojet} + \missET$ and $\text{soft-dilepton} + \text{jets} + \missET$ final states.

We will investigate the constraints from current LHC searches.
Moreover, the international high energy physics community is proposing future $pp$ collider projects, such as the Super Proton-Proton Collider (SPPC) at $\sqrt{s} \sim 70\text{--}100~\si{TeV}$~\cite{CEPC-SPPCStudyGroup:2015csa} and the $pp$ Future Circular Collider at $\sqrt{s} \sim 100~\si{TeV}$~\cite{Abada:2019lih}.
We will explore the prospect at a future $100~\si{TeV}$ $pp$ collider based on simulation.
Furthermore, electroweak interactions of the dark sector scalars also affect the electroweak oblique parameters at loop level, which can be examined in the electroweak precision test (EWPT).
Future $e^+ e^-$ collider projects, such as the Circular Electron-Positron Collider (CEPC)~\cite{CEPCStudyGroup:2018ghi}, the $e^+ e^-$ Future Circular Collider~\cite{Abada:2019zxq}, and the International Linear Collider~\cite{Baer:2013cma}, have plans to precisely measure the properties of the $Z$, $W$, and Higgs bosons and the top quark.
These measurements would greatly improve the determination of the electroweak oblique parameters~\cite{Fan:2014vta}.
We will estimate the corresponding sensitivity.

The paper is organized as follows. In Sec.~\ref{sec:model}, we introduce the inert sextuplet scalar model. In Sec~\ref{sec:split_decay}, we discuss the mass spectrum and decay processes. In Sec.~\ref{sec:lhc}, we investigate the current constraints from the $\text{monojet} + \missET$ and $\text{soft-dilepton} + \text{jets} + \missET$ searches at the LHC and the sensitivity at a 100~TeV $pp$ collider. In Sec.~\ref{sec:obliquepara}, we explore the current EWPT constraint and the future CEPC sensitivity.
In Sec.~\ref{sec:conslusion}, we give the conclusions and discussions.

\section{Inert sextuplet scalar model}
\label{sec:model}

In the inert sextuplet scalar model, the SM is extended with a $\SUtwoL$ sextuplet scalar field $S$ of hypercharge $Y=1/2$~\cite{Earl:2013fpa,Logan:2016ivc}, which can be expressed as
\begin{eqnarray}
S &=& \left(S^{3+},S^{2+},S^{+},S^{0},S^{-},S^{2-}\right)^\mathrm{T}
\nonumber\\
&=& \left(S^{11111},\sqrt{5}S^{11112},\sqrt{10}S^{11122},\sqrt{10}S^{11222},\sqrt{5}S^{12222},S^{22222}\right)^\mathrm{T}.
\end{eqnarray}
In the first line, the $S$ components are labeled by their electric charges. Note that $S$ must not be self-conjugated, implying $S^{-} \neq (S^{+})^\dag$ and $S^{2-} \neq (S^{2+})^\dag$.
In the second line, we reexpress the components with the tensor notation, where the sextuplet is represented by a totally symmetric rank-5 $\SUtwoL$ tensor $S^{ijklm}$
with normalization factors ensuring
\begin{equation}
S^\dag_{ijklm}S^{ijklm} = |{S^{3 + }}{|^2} + |{S^{2 + }}{|^2} + |{S^ + }{|^2} + |{S^0}{|^2} + |{S^ - }{|^2} + |{S^{2 - }}{|^2}.
\end{equation}
The electrically neutral component $S^{0}$ can be divided into two real scalars $\phi^0$ and $a^0$,
\begin{equation}
S^{0}=\frac{1}{\sqrt{2}}\left(\phi^{0}+ia^{0}\right).
\end{equation}
The lighter real scalar is a DM candidate.
For ensuring its stability, we impose a $Z_2$ symmetry $S\to -S$ to make the sextuplet inert.

The Lagrangian involving the sextuplet scalar reads
\begin{equation}\label{eq:L}
\mathcal{L}=\left(D^{\mu}S\right)^{\dag}_{ijklm} (D_{\mu}S)^{ijklm} - V(S).
\end{equation}
The covariant derivative of $S$ is given by
\begin{equation}\label{eq:CD}
(D_{\mu} S)^{ijklm} = \partial_\mu S^{ijklm}
- 5 i g (W_\mu)^i_n S^{njklm} - \frac{i}{2} g' B_\mu S^{ijklm}.
\end{equation}
Here $B_\mu$ is the $\UoneY$ gauge field and $(W_\mu)^i_j \equiv W_\mu^a (\sigma^a)_{ij} /2$ is the $\SUtwoL$ gauge fields understood as a $(1,1)$ tensor for the $\SUtwoL$ group, where $\sigma^a$ denote the Pauli matrices.
Thus, we have $(W_\mu)^1_1 = W^3_\mu/2 = - (W_\mu)^2_2$, $(W_\mu)^1_2 = W^+_\mu/\sqrt{2}$, and $(W_\mu)^2_1 = W^-_\mu/\sqrt{2}$.
The electroweak gauge couplings of the sextuplet components are given in Appendix~\ref{app:gauge}.

We write down the $Z_2$-invariant scalar potential involving $S$ as
\begin{eqnarray}
V(S)&=&m_{S}^{2}S^\dag_{ijklm}S^{ijklm}
+\lambda_{1}H^{\dag}_{i} S^{iklmn} S^{\dag}_{jklmn} H^{j}
+\lambda_{2}H^{\dag}_{i} S^{\dag}_{kmnpq} S^{lmnpq} H^{j}\epsilon^{ik}\epsilon_{lj}
\nonumber\\
&& ~ 
 +(\lambda_{3} H^{\dag}_{i}H^{\dag}_{j}S^{iklmn}S^{jpqrs}\epsilon_{kp}\epsilon_{lq}\epsilon_{mr}\epsilon_{ns}+\mathrm{H.c.})
\nonumber\\
&&~ + \text{quartic self-interaction terms of $S$},
\label{eq:potential}
\end{eqnarray}
where $H^i$ is the SM Higgs doublet, and the rank-2 Levi-Civita symbol satisfies $\epsilon^{ij} = (i\sigma^2)_{ij} = - \epsilon_{ij}$.
Note that $H_i^\dag {H^i}S_{jklmn}^\dag {S^{jklmn}} = H_i^\dag S_{jlkmn}^\dag {S^{ilkmn}}{H^j} + H_i^\dag S_{kmnpq}^\dag {S^{lmnpq}}{H^j}{\epsilon ^{ik}}{\epsilon _{lj}}$, and hence a $H_i^\dag {H^i}S_{jklmn}^\dag {S^{jklmn}}$ term is not independent.
As the quartic self-interaction terms of the sextuplet have no effects on the following discussions, we have not given their explicit expressions.
If $\lambda_3$ is complex, say, $\lambda_3 = |\lambda_3| e^{i\alpha}$, we can always absorb the phase $\alpha$ by redefining the sextuplet scalar field as $e^{i\alpha/2}S$ to make $\lambda_3$ real and positive.
Without loss of generality, we use $\lambda_3 \geq 0$ hereafter.

If the $Z_2$ symmetry $S\to -S$ is not respected, one may write down quartic terms like $H_i^\dag {S^{ijklm}}S_{jknop}^\dag {S^{nopqr}}{\epsilon _{lq}}{\epsilon _{mr}}$, which would lead to loop-induced decays of the DM candidate~\cite{Earl:2013fpa,Kumericki:2012bf,DiLuzio:2015oha}.
Therefore, such a $Z_2$ symmetry is not accidental, and we impose it by hand to stabilize the DM candidate.

After spontaneous breaking of the electroweak gauge symmetry, the $\lambda_1$, $\lambda_2$, and $\lambda_3$ terms in the unitary gauge become
\begin{eqnarray}
V(S) &\supset& \frac{{{\lambda _ + }}}{4}{(v + h)^2}(|{S^{3 + }}{|^2} + |{S^{2 + }}{|^2} + |{S^ + }{|^2} + |{S^0}{|^2} + |{S^ - }{|^2} + |{S^{2 - }}{|^2})
\nonumber\\
&&  + \frac{{{\lambda _ - }}}{{20}}{(v + h)^2}( - 5|{S^{3 + }}{|^2} - 3|{S^{2 + }}{|^2} - |{S^ + }{|^2} + |{S^0}{|^2} + 3|{S^ - }{|^2} + 5|{S^{2 - }}{|^2})
\nonumber\\
&& +\frac{{{\lambda _3}}}{{10}}{(v + h)^2}\Big[2\sqrt 5 {S^{2 - }}{S^{2 + }} - 4\sqrt 2 {S^ - }{S^ + } + 3{({S^0})^2} + \text{H.c.}\Big],
\end{eqnarray}
where $\lambda_\pm \equiv \lambda_1 \pm \lambda_2$ and the VEV of the Higgs doublet is $v = (\sqrt{2} G_\mathrm{F})^{-1/2}$.
Thus, the Higgs VEV contributes to the masses of the sextuplet components.
The mass terms read
\begin{eqnarray}
{\mathcal{L}_{{\mathrm{mass}}}} &=&  - m_{{S^{3 \pm }}}^2 |{S^{3 + }}{|^2} - \begin{pmatrix}
   {{{({S^{2 + }})}^\dag }} & {{S^{2 - }}}  \\
 \end{pmatrix}M_{2 + }^2\begin{pmatrix}
   {{S^{2 + }}}  \\
   {{{({S^{2 - }})}^\dag }}  \\
 \end{pmatrix} - \begin{pmatrix}
   {{{({S^ + })}^\dag}} & {{S^ - }}  \\
 \end{pmatrix}M_ + ^2\begin{pmatrix}
   {{S^ + }}  \\
   {{{({S^ - })}^\dag }}  \\
 \end{pmatrix}
\nonumber\\
&& - \frac{1}{2}m_{\phi^0} ^2{({\phi ^0})^2} - \frac{1}{2}m_{a^0}^2{({a^0})^2}.
\end{eqnarray}
where the masses of the triply charged scalar $S^{3\pm}$ and the neutral scalars $\phi^0$ and $a^0$ are given by
\begin{eqnarray}
m_{{S^{3 \pm }}}^2 &=& m_S^2 + \frac{1}{4}{\lambda _ + }{v^2} - \frac{1}{4}{\lambda _ - }{v^2},\\
m_{\phi^0} ^2 &=& m_S^2 + \frac{1}{4}{\lambda _ + }{v^2} + \frac{1}{{20}}{\lambda _ - }{v^2} + \frac{3}{5}{\lambda _3}{v^2},\\
m_{a^0}^2 &=& m_S^2 + \frac{1}{4}{\lambda _ + }{v^2} + \frac{1}{{20}}{\lambda _ - }{v^2} - \frac{3}{5}{\lambda _3}{v^2},
\end{eqnarray}
and the mass-squared matrices of the doubly and singly charged scalars are
\begin{eqnarray}
M_{2 + }^2 &=& \begin{pmatrix}
   m_S^2 + \dfrac{1}{4}{\lambda _ + }{v^2} - \dfrac{3}{{20}}{\lambda _ - }{v^2} & {\dfrac{{\sqrt 5 }}{5}{\lambda _3}{v^2}}  \\
   {\dfrac{{\sqrt 5 }}{5}{\lambda _3}{v^2}} & m_S^2 + \dfrac{1}{4}{\lambda _ + }{v^2} + \dfrac{1}{4}{\lambda _ - }{v^2}  \\
 \end{pmatrix},\\
M_ + ^2 &=& \begin{pmatrix}
   {m_S^2 + \dfrac{1}{4}{\lambda _ + }{v^2} - \dfrac{1}{{20}}{\lambda _ - }{v^2}} & { - \dfrac{{2\sqrt 2 }}{5}{\lambda _3}{v^2}}  \\
   { - \dfrac{{2\sqrt 2 }}{5}{\lambda _3}{v^2}} & {m_S^2 + \dfrac{1}{4}{\lambda _ + }{v^2} + \dfrac{3}{{20}}{\lambda _ - }{v^2}}  \\
 \end{pmatrix}.
\end{eqnarray}

The gauge and mass eigenstates of the doubly and singly charged scalars are connected by rotations,
\begin{eqnarray}
\left( {\begin{array}{*{20}{c}}
   {{S^{2 + }}}  \\
   {{{({S^{2 - }})}^\dag }}  \\
 \end{array} } \right) &=& R({\theta _{2 + }})\left( {\begin{array}{*{20}{c}}
   {S_1^{2 + }}  \\
   {S_2^{2 + }}  \\
 \end{array} } \right),\\
\left( {\begin{array}{*{20}{c}}
   {{S^ + }}  \\
   {{{({S^ - })}^\dag }}  \\
 \end{array} } \right) &=& R({\theta _ + })\left( {\begin{array}{*{20}{c}}
   {S_1^ + }  \\
   {S_2^ + }  \\
 \end{array} } \right),
\end{eqnarray}
with
\begin{equation}
R(\theta ) = \left( {\begin{array}{*{20}{c}}
   {\cos \theta } & { - \sin \theta }  \\
   {\sin \theta } & {\cos \theta }  \\
 \end{array} } \right).
\end{equation}
These rotations diagonalize the mass-squared matrices, resulting in masses given by
\begin{eqnarray}
m_{S_{1,2}^{2 \pm }}^2 &=& m_S^2 + \frac{1}{4}{\lambda _ + }{v^2} + \frac{1}{{20}}{\lambda _ - }{v^2} \mp \frac{1}{5}\sqrt {\lambda _ - ^2 + 5\lambda _3^2}\, {v^2},\\
m_{S_{1,2}^ \pm }^2 &=& m_S^2 + \frac{1}{4}{\lambda _ + }{v^2} + \frac{1}{{20}}{\lambda _ - }{v^2} \mp \frac{1}{{10}}\sqrt {\lambda _ - ^2 + 32\lambda _3^2}\, {v^2}.
\end{eqnarray}
Here we adopt a mass hierarchy convention of $m_{S_1^{2 \pm }}^2 \leq m_{S_2^{2 \pm }}^2$ and $m_{S_1^ \pm }^2 \leq m_{S_2^ \pm }^2$.
The rotation angles satisfy
\begin{eqnarray}
\sin {\theta _{2 + }} &=& \frac{{ - \sqrt {5/2}\, {\lambda _3}}}{{\sqrt {\lambda _ - ^2 + 5\lambda _3^2 + {\lambda _ - }\sqrt {\lambda _ - ^2 + 5\lambda _3^2} } }},\\
\sin {\theta _ + } &=& \frac{{4{\lambda _3}}}{{\sqrt {\lambda _ - ^2 + 32\lambda _3^2 + {\lambda _ - }\sqrt {\lambda _ - ^2 + 32\lambda _3^2} } }}.
\end{eqnarray}

Note that $\lambda_+$ equally contributes to all the scalar masses, and mass splittings are only induced by $\lambda_-$ and $\lambda_3$.
Since we have adopted $\lambda_3 \geq 0$, $a^0$ is not heavier than $\phi^0$.
In order to prevent any charged scalar lighter than $a^0$, we should require $|\lambda_-| \leq 2\lambda_3$.
In this case, $a^0$ is a viable DM candidate that cannot decay.
The mass hierarchy of the dark sector scalars is
\begin{equation}
m_{a^0}\leq m_{S^{\pm}_{1}}\leq m_{S^{2\pm}_{1}}\leq m_{S^{3\pm}}\leq m_{S^{2\pm}_{2}}\leq m_{S^{\pm}_{2}}\leq m_{\phi^0}.
\end{equation}
Free parameters of the model are adopted to be
\begin{equation}
m_S,~~\lambda_+,~~\lambda_-,~~\lambda_3.
\end{equation}

The trilinear coupling between the Higgs boson $h$ and the DM candidate $a^0$ is given by
\begin{eqnarray}
\mathcal{L} &\supset&
 \frac{1}{2}\lambda_{haa}v\,ha^{0}a^{0},\\
\lambda_{haa} &=& -\frac{1}{2}\lambda_{+}-\frac{1}{10}\lambda_{-}+\frac{6}{5}\lambda_{3}.
\end{eqnarray}
This coupling induces spin-independent (SI) $a^0$ scattering off nucleons in DM direct detection experiments.
Making use of dimension-5 effective operators~\cite{Yu:2011by}, we derive the $a^0$-nucleon SI  scattering cross section as
\begin{equation}
\sigma _{{a^0}N}^{{\mathrm{SI}}} = \frac{{\lambda _{haa}^2m_N^4{{[2 + 7(f_u^N + f_d^N + f_s^N)]}^2}}}{{324\pi m_h^4{{({m_{{a^0}}} + {m_N})}^2}}},\quad N = p,n,
\end{equation}
where the form factors of nucleons $f^N_{u,d,s}$ can be found in Ref.~\cite{Ellis:2000ds}.

Note that the condition
\begin{equation}\label{eq:flat_cond}
-\frac{1}{2}\lambda_{+}-\frac{1}{10}\lambda_{-}+\frac{6}{5}\lambda_{3} = 0
\end{equation}
corresponds to flat directions among the scalar couplings, where the Higgs VEV gives no contribution to the  $a^0$ mass and the $ha^0a^0$ coupling vanishes.
In this case, $a^0$-nucleon scattering is absent at tree level, and direct detection experiments would hardly constrain the model.

\section{Mass splittings and scalar decays}
\label{sec:split_decay}

\begin{figure}[!t]
\centering
\subfigure[~Mass splittings\label{fig:masssplit}]
{\includegraphics[width=0.49\textwidth]{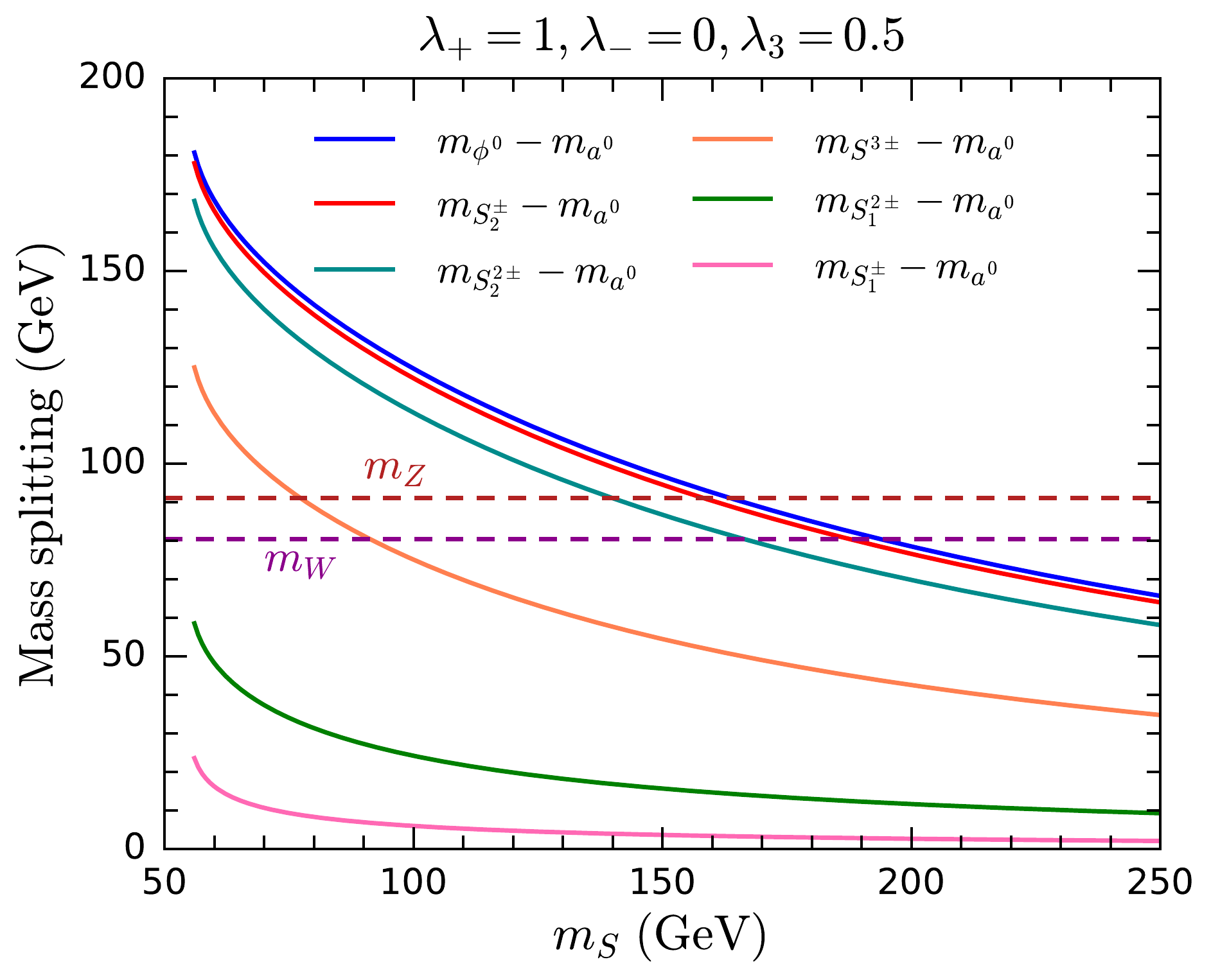}}
\subfigure[~Contours of $m_{\phi^0} - m_{a^0}$\label{fig:split:lam3}]
{\includegraphics[width=0.49\textwidth]{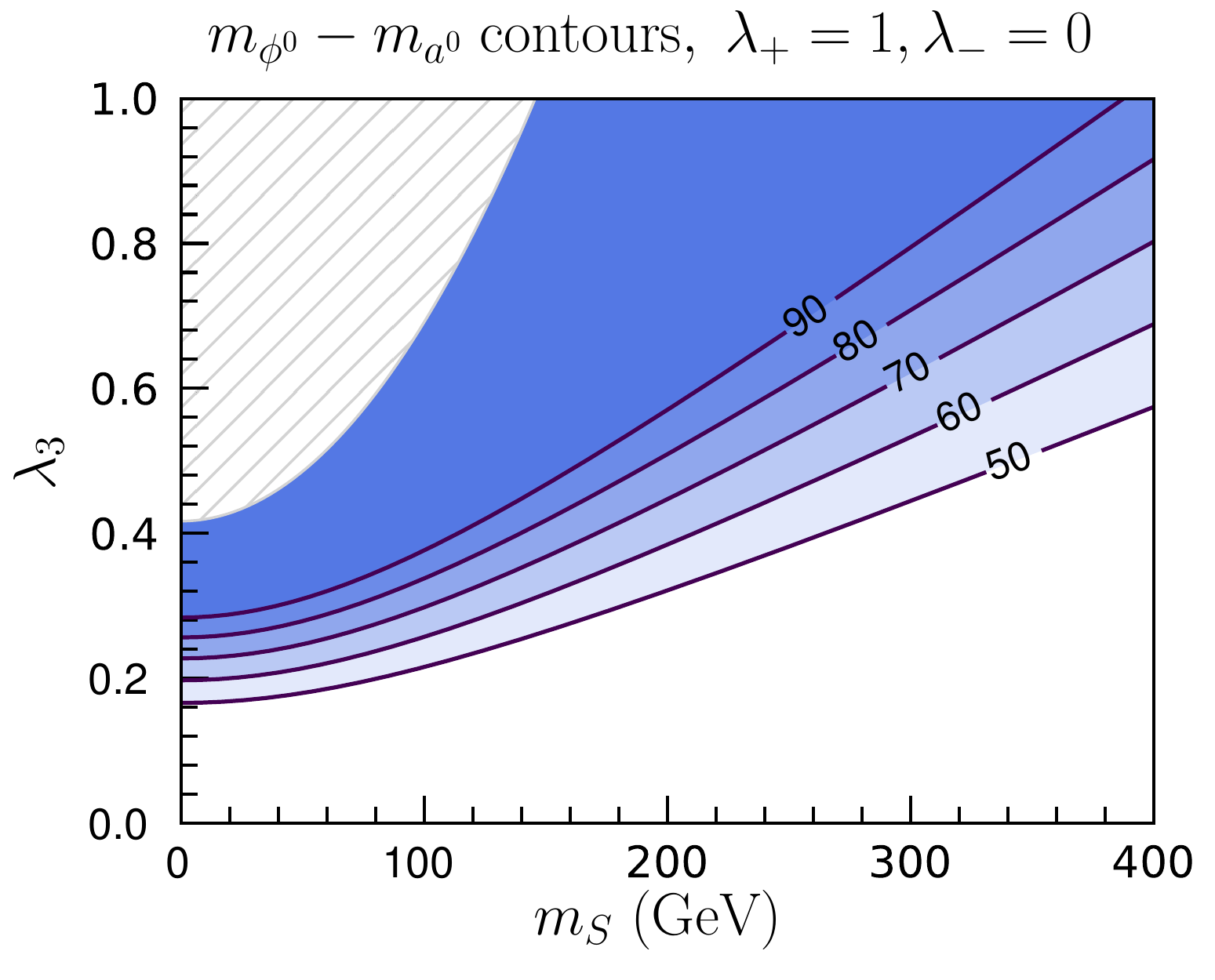}}
\caption{Mass splittings among the dark sector scalars.
(a) Mass splittings between $a^0$ and the other scalar bosons for $\lambda_+ = 1$, $\lambda_-=0$, and $\lambda_3 = 0.5$.
(b) Contours of the mass splitting in GeV between $\phi^0$ and $a^0$ in the $m_S$-$\lambda_3$ plane for $\lambda_+ = 1$ and $\lambda_- = 0$.}
\label{fig:split}
\end{figure}

If electroweak gauge symmetry is strictly respected, the components of the sextuplet scalar must have a common mass $m_S$.
Nonetheless, the VEV of the Higgs field breaks the degeneracy, leading nonzero mass splittings among the dark sector scalars.
Figure~\ref{fig:masssplit} shows the mass splittings between the DM candidate $a^0$ and the other scalar bosons as functions of $m_S$ for $\lambda_+ = 1$, $\lambda_-=0$, and $\lambda_3 = 0.5$.
Note that $m_{S}$ universally contributes to all the scalar masses.
As $m_{S}$ increases, the contributions to the masses from $\lambda_-$ and $\lambda_3$ become relatively small, and hence the mass splittings decrease.
For fixed $m_{S}$, smaller $\lambda_+$ means that the common VEV contributions to the scalar masses are smaller, while smaller $\lambda_-$ and $\lambda_3$ lead to smaller mass splittings.
In Fig.~\ref{fig:masssplit}, we also indicate the masses of the $W^\pm$ and $Z$ bosons to show whether $W/Z$ is on- or off-shell in the scalar decays.

The largest mass splitting among the scalars is $m_{\phi^0} - m_{a^0}$, whose contours in the $m_S$-$\lambda_3$ plane are demonstrated in Fig.~\ref{fig:split:lam3} for $\lambda_+ = 1$ and $\lambda_- = 0$.
We must ensure that the DM candidate $a^{0}$ has a physical mass, i.e., $m_a^2>0$, corresponding to 
$m_{S}^{2} > 3\lambda_3v^2/5 - \lambda_+v^2/4 - \lambda_-v^2/20$.
The hatched region in Fig.~\ref{fig:split:lam3} is ruled out because it violates this condition.
For $m_S \gtrsim 300~\si{GeV}$, the mass spectrum is rather compressed.

\begin{figure}[!t]
\centering
\subfigure[~$S^{3+}/S^{2+}_i/S^+_i \to W^{+(*)} + S^{2+}_1/S^+_1/a^0$]
{\includegraphics[height=.2\textwidth]{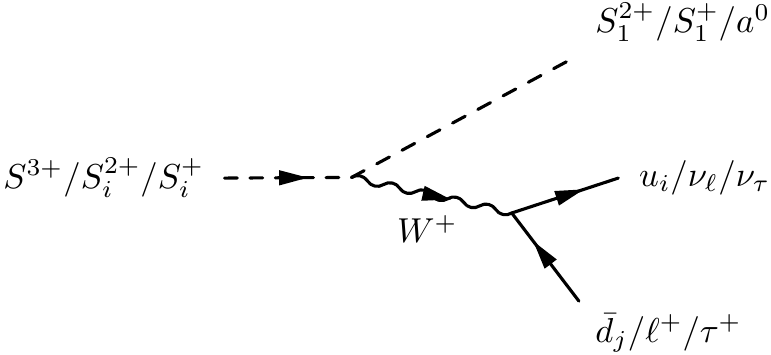}}
\hspace{2em}
\subfigure[~$S_2^{2+}/S_2^+/\phi^0 \to W^{-(*)} + S^{3+}/S_i^{2+}/S_i^+$]
{\includegraphics[height=.2\textwidth]{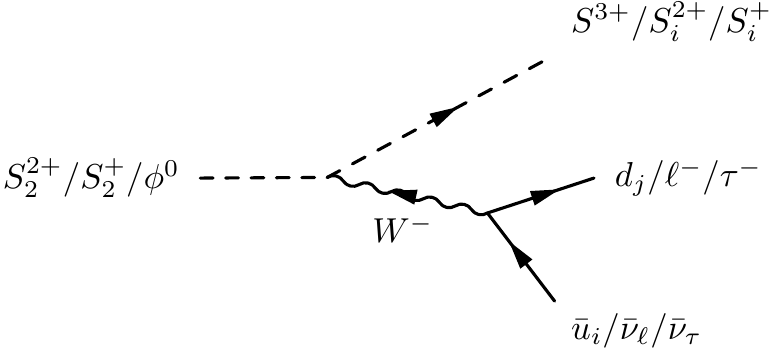}}
\subfigure[~$S_2^{2+}/S_2^+/\phi^0 \to Z^{(*)} + S_1^{2+}/S_1^+/a^0$\label{fig:decay:Z}]
{\includegraphics[height=.2\textwidth]{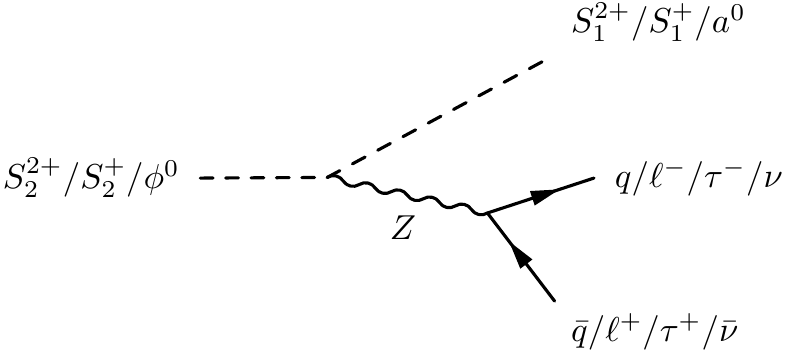}}
\hspace{2em}
\subfigure[~$S_2^{2+}/S_2^+ \to h^{(*)} + S_1^{2+}/S_1^+$]
{\includegraphics[height=.2\textwidth]{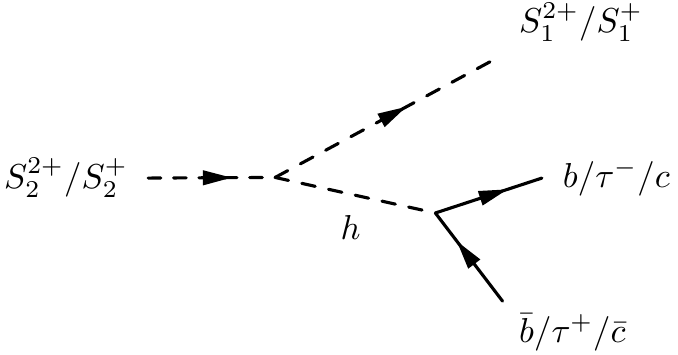}}
\caption{Typical 3-body decay diagrams for $S^{3+}/S^{2+}_i/S^+_i \to W^{+(*)} + S^{2+}_1/S^+_1/a^0$~(a), $S_2^{2+}/S_2^+/\phi^0 \to W^{-(*)} + S^{3+}/S_i^{2+}/S_i^+$~(b), $S_2^{2+}/S_2^+/\phi^0 \to Z^{(*)} + S_1^{2+}/S_1^+/a^0$~(c), and $S_2^{2+}/S_2^+ \to h^{(*)} + S_1^{2+}/S_1^+$~(d).
Here $u_i = (u,c)$, $d_i = (d,s)$, $\ell = (e,\mu)$, and $q=(d,u,s,c,b)$.}
\label{fig:decay}
\end{figure}

Dark sector scalars heavier than $a^0$ are unstable.
They can decay into lighter states through the mediation of the weak gauge bosons $W^\pm$ and $Z$ and the Higgs boson $h$.
The mediation particles could be either on or off shell, depending on the mass splittings.
Important decay channels include
$S^{3\pm}/S_i^{2\pm}/S_i^\pm \to S_1^{2\pm}/S_1^{\pm}/a^0 + W^{\pm(*)}$,
$S_2^{2\pm}/S_2^\pm/\phi^0 \to S^{3\pm}/S_i^{2\pm}/S_i^{\pm} + W^{\mp(*)}$,
$S_2^{2\pm}/S_2^{\pm}/\phi^0 \to S_1^{2\pm}/S_1^{\pm}/a^0 + Z^{(*)}$,
and $S_2^{2\pm}/S_2^{\pm} \to S_1^{2\pm}/S_1^{\pm} + h^{(*)}$.
The gauge and Higgs bosons subsequently convert to lighter SM particles.
Typical 3-body decay diagrams are presented in Fig.~\ref{fig:decay}.
Because of the $Z_2$ symmetry, all decay chains finally end at the DM candidate $a^0$.
Typical mass splittings in this model are not large enough for an on-shell Higgs boson, and Higgs mediated decays into SM fermions are highly suppressed by the Yukawa couplings.
Therefore, Higgs induced decays are commonly negligible.

For a scalar boson with 4-momentum $p^\mu$ and mass $m$ decaying into three particles with 4-momenta $k_1^\mu$, $k_2^\mu$, and $k_3^\mu$ and masses $m_1$, $m_2$, and $m_3$, the partial decay width can be expressed as~\cite{Tanabashi:2018oca}
\begin{equation}
\Gamma = \frac{1}{256\pi^3 m^{3}}\int_{s_{12}^{\mathrm{min}}}^{s_{12}^{\mathrm{max}}}ds_{12}\int_{s_{23}^{\mathrm{min}}}^{s_{23}^{\mathrm{max}}}ds_{23}\,|\mathcal{M}|^{2},
\end{equation}
where $|\mathcal{M}|^{2}$ is the invariant amplitude squared with summation over final state spins.
The lower and upper limits of $s_{23}\equiv(k_{2}+k_{3})^{2}$ are given by
\begin{eqnarray}
s_{23}^{\mathrm{min}}&=&(\tilde{E}_{2}+\tilde{E}_{3})^{2}-\Big(\sqrt{\tilde{E}_{2}^{2}-m_{2}^{2}}+\sqrt{\tilde{E}_{3}^{2}-m_{3}^{2}}\Big)^{2},\\ s_{23}^{\mathrm{max}}&=&(\tilde{E}_{2}+\tilde{E}_{3})^{2}-\Big(\sqrt{\tilde{E}_{2}^{2}-m_{2}^{2}}-\sqrt{\tilde{E}_{3}^{2}-m_{3}^{2}}\Big)^{2},
\end{eqnarray}
with 
\begin{eqnarray}
\tilde{E}_{2}&\equiv&\frac{s_{12}-m_{1}^{2}+m_{2}^{2}}{2\sqrt{s_{12}}},\\
\tilde{E}_{3}&\equiv&\frac{m^{2}-s_{12}-m_{3}^{2}}{2\sqrt{s_{12}}}.
\end{eqnarray}
The lower and upper limits of $s_{12}\equiv(k_{1}+k_{2})^{2}$ are $s_{12}^{\mathrm{min}}=(m_{1}+m_{2})^{2}$ and $s_{12}^{\mathrm{max}}=(m-m_{3})^{2}$.

For decay processes of the dark sector scalar bosons, we derive the tree-level $|\mathcal{M}|^{2}$ and calculate the partial widths and branching ratios utilizing the formulas above.
For instance, the amplitude squared for the $Z$-mediated decay channel $\phi^{0}(p)\rightarrow a^{0}(k_{1})+\bar{f}(k_{2})+f(k_{3})$, where the 4-momenta are indicated between the parentheses, is obtained as
\begin{eqnarray}
|\mathcal{M}|^{2}
&=&\frac{g^{4}}{2c_\mathrm{W}^{4}[(s_{23}-m_{Z}^{2})^{2}+m_{Z}^{2}\Gamma_{Z}^{2}]}
\nonumber\\
&&\times\Big\{\big(g_\mathrm{V}^{f}\big)^{2}\big[(s_{12}-m_{f}^{2})(m_{\phi^0}^{2}+m_{a^0}^{2}+m_{f}^{2}-s_{12}-s_{23})-m_{\phi^0}^{2}m_{a^0}^{2}\big]
\nonumber\\
&&\quad~+\big(g_\mathrm{A}^{f}\big)^{2}\big[s_{12}(m_{\phi^0}^{2}+m_{a^0}^{2}-s_{12}-s_{23})-m_{\phi^0}^{2}m_{a^0}^{2}+m_{f}^{2}(m_{\phi^0}^{2}+m_{a^0}^{2}+2s_{12}-m_{f}^{2})
\nonumber\\
&&\qquad\qquad\quad~+{m_{f}^{2}}{m_{Z}^{-4}}(m_{\phi^0}^{2}-m_{a^0}^{2})^{2}(s_{23}-2m_{Z}^{2})\big]\Big\},
\end{eqnarray}
Here $c_\mathrm{W}\equiv \cos \theta_\mathrm{W}$ with $\theta_\mathrm{W} = \tan^{-1} (g'/g)$ denoting the weak mixing angle.
$\Gamma_Z$ denotes the total width of $Z$, and $g_\mathrm{V}^{f}$ and $g_\mathrm{A}^{f}$ are the vector and axial-vector coupling coefficients of the SM fermion $f$ to $Z$, respectively.
On the other hand, the amplitude squared for the $W$-mediated decay channel $\phi^{0}(p)\rightarrow S_{1}^{+}(k_{1})+\bar{d}_j(k_{2})+u_i(k_{3})$ with $u_i = (u,c)$ and $d_i = (d,s)$ is given by
\begin{eqnarray}
|\mathcal{M}{|^2} &=& \frac{{g^4}|{V_{ij}}|^2{{(3\cos {\theta _+} - 2\sqrt 2 \sin {\theta_+ })}^2}}{{8[{{({s_{23}} - m_W^2)}^2} + m_W^2\Gamma _W^2]}}
\nonumber\\
&& \times\Big\{ 4(m_{{\phi ^0}}^2 - {s_{12}})({s_{12}} - m_{S_1^ \pm }^2)+ (4{s_{12}} + {s_{23}})(m_{{u_i}}^2 + m_{{d_j}}^2) - 4{s_{12}}{s_{23}} - {(m_{{u_i}}^2 + m_{{d_j}}^2)^2}
\nonumber\\
&& - 2m_W^{ - 2}(m_{{\phi ^0}}^2 - m_{S_1^ \pm }^2)\big[(2{s_{12}} + {s_{23}})(m_{{u_i}}^2 - m_{{d_j}}^2) - 2m_{S_1^ \pm }^2m_{{u_i}}^2 + 2m_{\phi^0} ^2m_{{d_j}}^2 - m_{{u_i}}^4 + m_{{d_j}}^4\big]
\nonumber\\
&& - m_W^{ - 4}{(m_{{\phi ^0}}^2 - m_{S_1^ \pm }^2)^2}\big[{(m_{{u_i}}^2 - m_{{d_j}}^2)^2} - {s_{23}}(m_{{u_i}}^2 + m_{{d_j}}^2)\big]\Big\},
\end{eqnarray}
where $\Gamma_W$ is the total width of $W$ and $V_{ij}$ is the Cabibbo-Kobayashi-Maskawa matrix.

\begin{figure}[!t]
\centering
\subfigure[~$S_1^+$ decay channels\label{fig:decay_BR:S+1}]
{\includegraphics[width=0.49\textwidth]{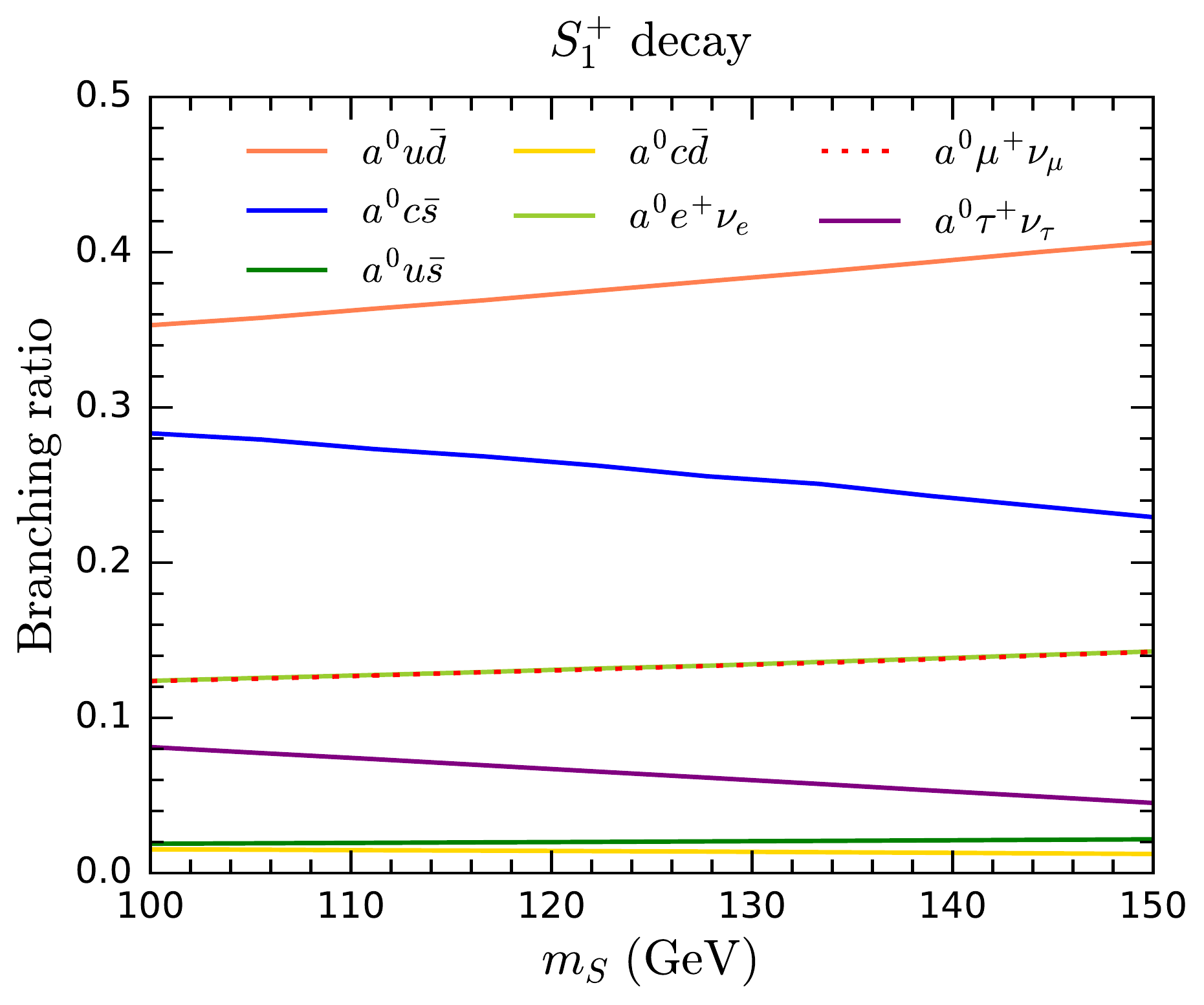}}
\subfigure[~$\phi^0$ decay channels\label{fig:decay:phi}]
{\includegraphics[width=0.49\textwidth]{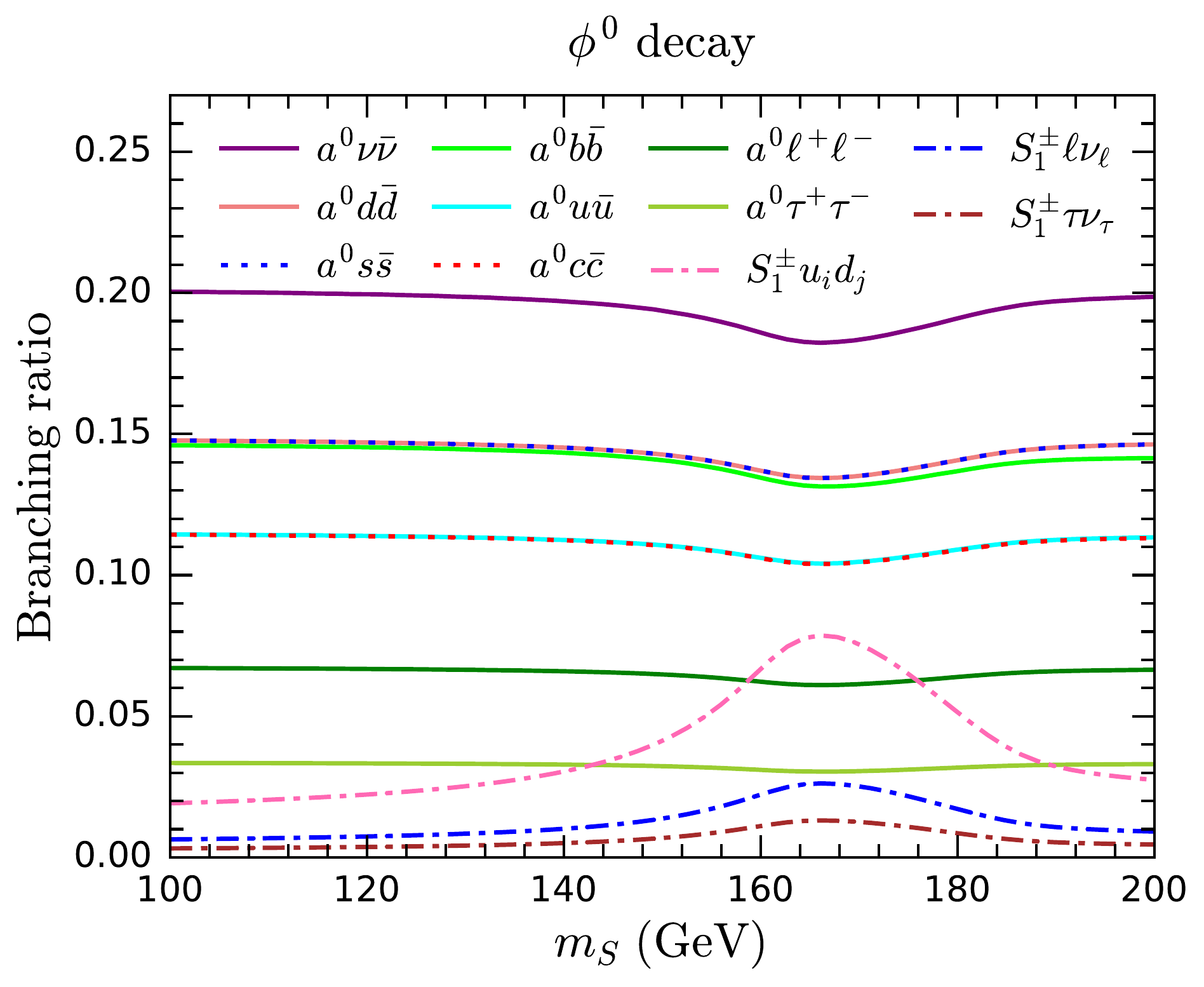}}
\caption{Branching ratios of $S_1^+$ (a) and $\phi^0$ (b) decay channels for $\lambda_{+}=1$, $\lambda_{-}=0$, and $\lambda_{3}=0.5$.}
\label{fig:decay_BR}
\end{figure}

In Figs.~\ref{fig:decay_BR:S+1} and \ref{fig:decay:phi}, we show the branching ratios of the decay channels of $S_1^+$ and $\phi^{0}$, respectively.
The parameters are fixed to be $\lambda_{+}=1$, $\lambda_{-}=0$, and $\lambda_{3}=0.5$, as the same as in Fig.~\ref{fig:masssplit}.

The $S_1^+$ boson only has $W$-mediated decay channels.
For the parameters adopted in Fig.~\ref{fig:decay_BR:S+1}, the $W$ boson is always off shell.
In the $S_1^+$ decay channels, the $S_1^+\to a^0 u\bar{d}$ branching ratio $\mathcal{B}(S_1^+\to a^0 u\bar{d})$ is the largest, while $\mathcal{B}(S_1^+\to a^0 c\bar{s})$ is the second largest due to the suppression by the $c$ and $s$ quark masses.
Because of the Cabibbo suppression, $\mathcal{B}(S_1^+\to a^0 u\bar{s})$ and $\mathcal{B}(S_1^+\to a^0 c\bar{d})$ are negligible.
$\mathcal{B}(S_1^+\to a^0 e^+\nu_e)$ and $\mathcal{B}(S_1^+\to a^0 \mu^+\nu_\mu)$ are basically identical, reflecting the lepton universality, while $\mathcal{B}(S_1^+\to a^0 \tau^+\nu_\tau)$ is smaller, suppressed by the $\tau$ lepton mass.
As $m_S$ increases, the mass splitting between $S_1^+$ and $a^0$ decreases, leading to smaller phase spaces in the final states.
Consequently, the $S_1^+\to a^0 c\bar{s}$ and $S_1^+\to a^0 \tau^+\nu_\tau$ channels are more suppressed for larger $m_S$, resulting in larger $\mathcal{B}(S_1^+\to a^0 u\bar{d})$ and $\mathcal{B}(S_1^+\to a^0 \ell^+\nu_\ell)$, where $\ell = e,\mu$.

Figure~\ref{fig:decay:phi} demonstrates the $Z$-mediated $\phi^0$ decay channels into $a^0$ as well as the $W$-mediated channels into $S_1^\pm$.
Because of the tiny mass splitting between $\phi^0$ and $S_2^\pm$, the $\phi^0$ decay channels into $S_2^\pm$ are negligible and not shown in the plot.
In the $Z$-mediated channels, $\mathcal{B}(\phi^0 \to a^0 \nu\bar\nu)$ is as large as $\sim 20\%$.
The $\phi^0$ decay into a down-type (up-type) quark pair has a branching ratio $\sim 15\%$ ($\sim 11\%$), while the $\phi^0$ decay into a charged lepton pair has a branching ratio $\sim 3\%$.

For a region around $m_S \sim 165~\si{GeV}$ in Fig.~\ref{fig:decay:phi}, the $Z$-mediated channels commonly decrease; in contrast, the $W$-mediated channels commonly increase.
This can be understood by comparing with the mass splittings shown in Fig.~\ref{fig:masssplit}.
For $m_S \lesssim 160~\si{GeV}$, we have $m_{\phi^0} - m_{a^0} \gtrsim m_Z$ and the on-shell $Z$ and $W$ bosons dominate over the phase spaces.
In this case, the $\phi^0$ decays can be treated as 2-body decays $\phi^0 \to a^0 Z$ and $\phi^0 \to S_1^\pm W^\mp$, where the weak gauge bosons subsequently decay into SM fermions.
For $m_S \gtrsim 160~\si{GeV}$, however, the $Z$ boson becomes off shell.
Thus, the $Z$-mediated channels are suppressed by the phase spaces of 3-body final states.
A similar suppression on the $W$-mediated channels happens later, at $m_S \gtrsim 185~\si{GeV}$, where the behaviors of the two types of channels are gradually restored.

\section{Searches at $pp$ colliders}
\label{sec:lhc}

High energy colliders are primary tools for finding new particles.
Currently, the highest center-of-mass energy in $pp$ collisions is $\sqrt{s} =13~\si{TeV}$, achieved by the LHC, which will be soon upgraded to $\sqrt{s} =14~\si{TeV}$.
Moreover, a future $pp$ collider with $\sqrt{s} \sim 100~\si{TeV}$ would provide great opportunities for discovering the new scalar bosons in the inert sextuplet scalar model.

\begin{figure}[!t]
\centering
\subfigure[~$W^+$ mediation]
{\includegraphics[height=.21\textwidth]{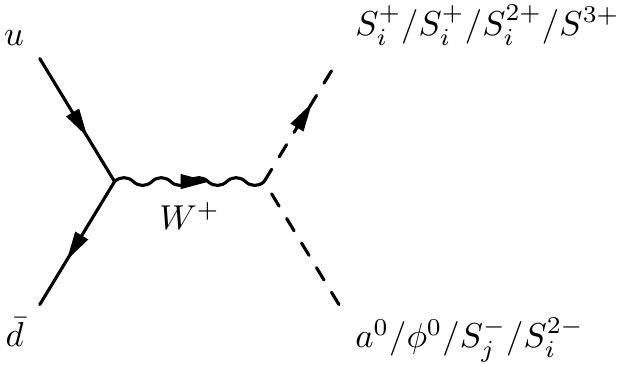}}
\hspace{1em}
\subfigure[~$W^-$ mediation]
{\includegraphics[height=.21\textwidth]{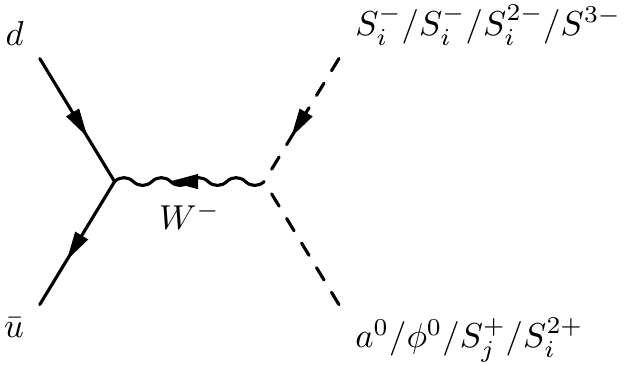}}
\subfigure[~$Z$ mediation]
{\includegraphics[height=.21\textwidth]{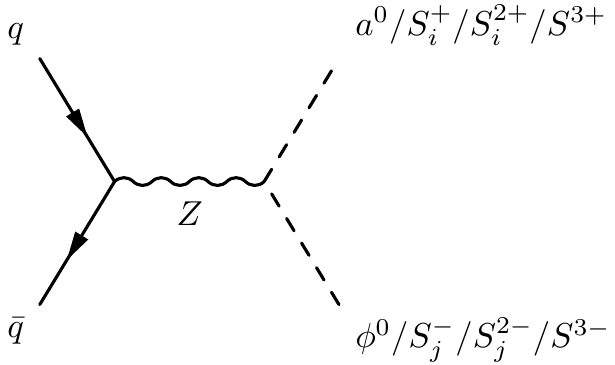}}
\hspace{1em}
\subfigure[~$\gamma$ mediation]
{\includegraphics[height=.21\textwidth]{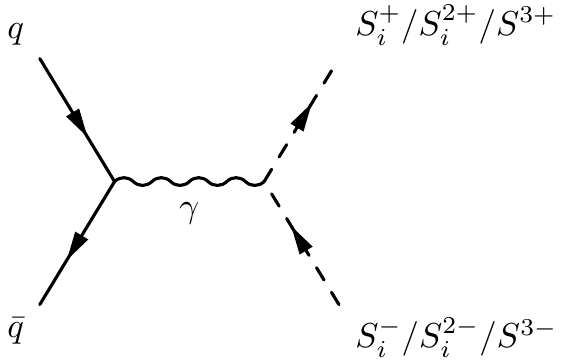}}
\caption{Typical Feynman diagrams for the scalar pair production at $pp$ colliders, including $W^+$-mediated $u + \bar{d} \to S^+_i/S^+_i/S^{2+}_i/S^{3+} + a^0/\phi^0/S^-_j/S^{2-}_i$~(a), $W^-$-mediated $d + \bar{u} \to S^-_i/S^-_i/S^{2-}_i/S^{3-} + a^0/\phi^0/S^+_j/S^{2+}_i$~(b), $Z$-mediated $q + \bar{q} \to a^0/S^+_i/S^{2+}_i/S^{3+} + \phi^0/S^-_j/S^{2-}_j/S^{3-}$~(c), and $\gamma$-mediated $q + \bar{q} \to S^+_i/S^{2+}_i/S^{3+} + S^-_i/S^{2-}_i/S^{3-}$~(d).}
\label{fig:prod}
\end{figure}

\begin{figure}[!t]
\centering
\includegraphics[width=0.5\textwidth]{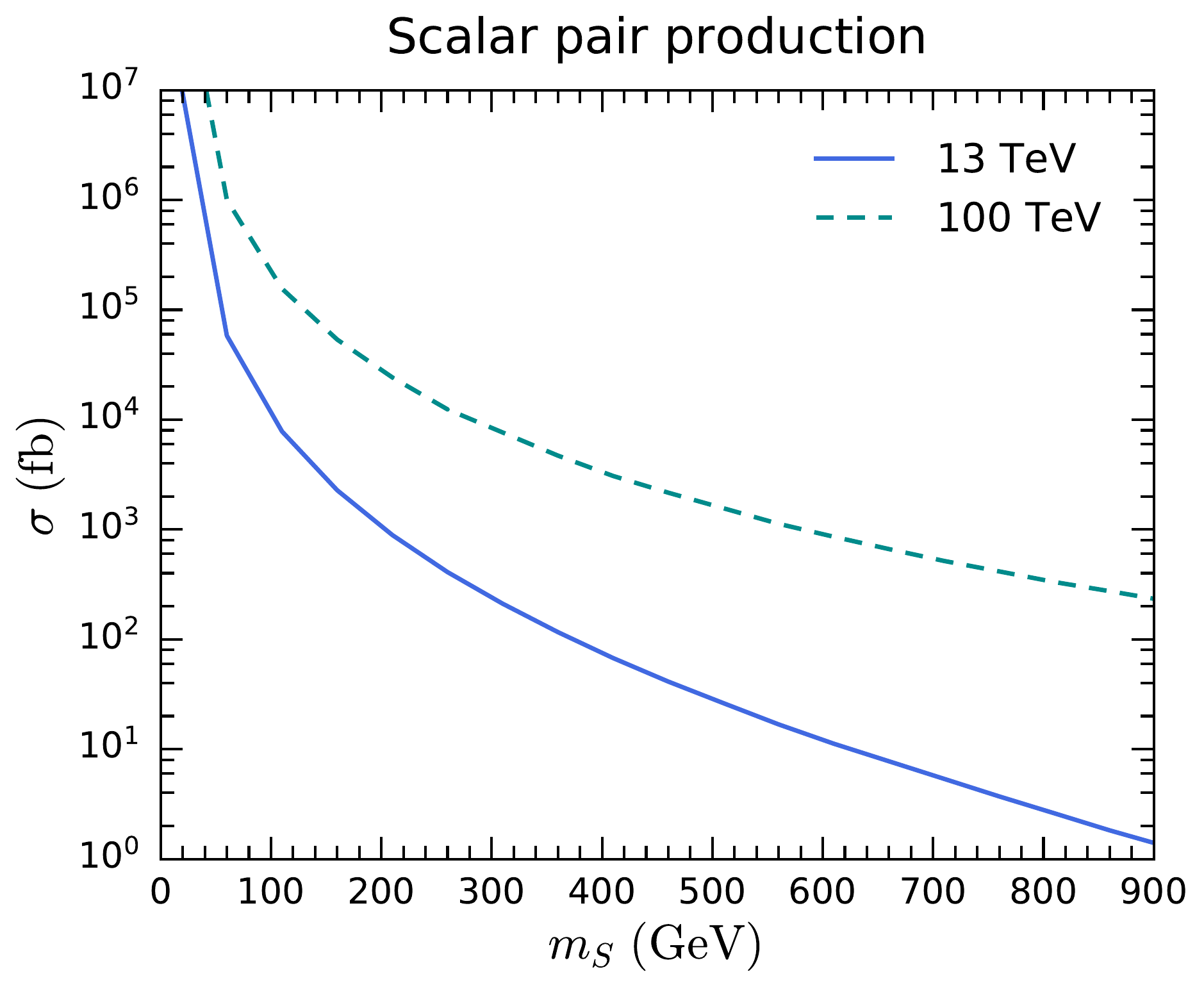}
\caption{Production cross sections as functions of $m_S$ for $pp \to S_i S_j + \text{jets}$ with $S_i = (a^0, \phi^0, S_1^\pm,S_2^\pm,S_1^{2\pm},S_2^{2\pm},S^{3\pm})$ at $\sqrt{s}=13~\si{TeV}$ and $100~\si{TeV}$.
The couplings are fixed as $\lambda_{+}=\lambda_{-}=\lambda_{3}=0$.}
\label{productionxsc}
\end{figure}

The dark sector scalars can be directly produced in pairs at $pp$ colliders through electroweak gauge interactions.
The inclusive production processes can be represented as $pp \to S_i S_j + \text{jets}$ with $S_i = (a^0, \phi^0, S_1^\pm,S_2^\pm,S_1^{2\pm},S_2^{2\pm},S^{3\pm})$.
The corresponding Feynman diagrams at parton level are demonstrated in Fig.~\ref{fig:prod}.
The inclusive production cross section as a function of $m_S$ is shown in Fig.~\ref{productionxsc} for $\lambda_{+}=\lambda_{-}=\lambda_{3}=0$, which implies that all dark sector scalars have a common mass $m_S$.
For $m_S \sim 200~\si{GeV}$, the cross section is $\sim 1~\si{pb}$ at $\sqrt{s} = 13~\si{TeV}$, and increases to $\sim 30~\si{pb}$ at $\sqrt{s} = 100~\si{TeV}$.
For $m_S \gtrsim 700~\si{GeV}$, the cross section increases by more than two orders of magnitude when $\sqrt{s}$ is promoted from $13~\si{TeV}$ to $100~\si{TeV}$.

Since the only stable scalar is the DM candidate $a^0$, all other scalars finally decay into $a^0$, associated with visible particles.
General detectors at the LHC and future $pp$ colliders cannot probe $a^0$, which would give rise to a missing transverse momentum $\misspT$ in the final state.
The missing transverse energy is defined as $\missET \equiv |\misspT|$, characterizing the energy amount of the particles that do not recorded by the detector.
Therefore, a large $\missET$ is an important imprint for $a^0$.
According to other information from the associating visible particles, we can utilize different channels to search for the dark sector scalars.
In the following subsections, we separately discuss the $\text{monojet} + \missET$ and $\text{soft-dilepton} + \text{jets} + \missET$ search channels.

We will investigate the current LHC constraints on the inert sextuplet scalar model by reinterpreting the experimental analyses at $\sqrt{s} = 13~\si{TeV}$ and study the future sensitivity at a 100~TeV $pp$ collider.
For this purpose, we need to generate Monte Carlo simulation samples for the signals and relevant backgrounds.
The inert sextuplet model is implemented by \texttt{FeynRules~2}~\cite{Alloul:2013bka}, which reads off the Lagrangian we defined and outputs the Feynman rules for Monte Carlo tools.
Moreover, \texttt{MadGraph5\_aMG@NLO~2}~\cite{Alwall:2014hca} is utilized to automatically generate Feynman diagrams, calculate matrix elements, perform phase space integrations, and generate parton-level event samples.
Furthermore, we use ~\texttt{PYTHIA~8}~\cite{Sjostrand:2014zea} to carry out parton shower, hardronization, and particle decays.
The MLM matching scheme~\cite{Mangano:2006rw} is adopted for merging matrix element and parton shower calculations.
Finally, \texttt{Delphes~3}~\cite{deFavereau:2013fsa} is used for a fast, simplified detector simulation with a parametrization of the ATLAS detector.

\subsection{$\text{Monojet}+\missET$ channel}

From the analysis in Sec.~\ref{sec:split_decay} as well as Fig.~\ref{fig:split}, we know that the mass splittings among the dark sector scalars are quite small for small scalar couplings and/or large $m_S$.
Therefore, in a large portion of the parameter space, the visible products from the scalar decays are quite soft.
It is possible to utilize soft leptons (electrons and muons) for probing the signal, and we will discuss this possibility in the next subsection.
In the present subsection, we choose to neglect the soft particles and consider at least one energetic jet originated from initial state radiation to recoil against the scalar pair for achieving a large $\missET$.
Isolated leptons are further vetoed to give a clear signature.
This leads to the $\text{monojet}+\missET$ search channel, which has been widely used at the LHC for DM searches~\cite{Beltran:2010ww,Rajaraman:2011wf,Fox:2011pm,Yu:2012kj,Xiang:2015lfa,Wang:2017sxx,Zeng:2019tlw}.

The primary background in the $\text{monojet}+\missET$ channel is $Z(\rightarrow\nu\bar{\nu})+\text{jets}$, where the neutrinos from $Z$ decays cannot be detected and also lead to missing transverse energy.
The subdominant background is $W(\rightarrow \ell\nu)+\text{jets}$, where the charged lepton $\ell$ could either fall outside of the detector acceptance or be combined into a jet~\cite{Beltran:2010ww}.
Minor backgrounds include small fractions of $Z/\gamma^{*}(\rightarrow \ell^{+}\ell^{-})+\text{jets}$, multijet, $t\bar{t}+\text{jets}$, single top, diboson ($WW,WZ,ZZ$) processes~\cite{Aaboud:2016tnv}.
The multijet background can be efficiently reduced by requiring a sufficient large azimuthal angle $\Delta\phi(j,\misspT)$ between $\misspT$ and each reconstructed jet $j$.

\subsubsection{LHC constraint}

Firstly, we utilize the $\text{monojet}+\missET$ analysis from the ATLAS experiment with an integrated luminosity of $36.1~\si{fb^{-1}}$ at $\sqrt{s}=13~\si{TeV}$~\cite{Aaboud:2017phn} to evaluate the monojet constraint on the model.
In the ATLAS analysis, reconstructed jets are required to have $\pT > 30~\si{GeV}$ and $|\eta|<2.8$, while reconstructed isolated electrons (muons) should have $\pT > 20~(10)~\si{GeV}$ and $|\eta|<2.47~(2.5)$.
Then the following selection cuts are applied for increasing the signal significance: (a) $\missET > 250~\si{GeV}$; (b) a leading jet with $\pT > 250~\si{GeV}$ and $|\eta|<2.4$; (c) no more than four jets; (d) $\Delta\phi(j,\misspT) > 0.4$; (e) no isolated electron or muon.

\begin{table}[!t]
\renewcommand{\arraystretch}{1.2}
\setlength{\tabcolsep}{.6em}
\centering
\caption{Signal regions in the ATLAS $\text{monojet}+\missET$ analysis at $\sqrt{s} = 13~\si{TeV}$~\cite{Aaboud:2017phn}.
$\sigma_\mathrm{vis}^{95}$ is the observed 95\% C.L. upper limit on the visible cross section $\sigma_\mathrm{vis}$.}
\label{tablesr}
\begin{tabular}{ccccccccc}
\hline\hline
Signal regions & IM1  & IM2   & IM3   & IM4  & IM5  & IM6   &IM7   & IM8   \\
\hline
$\missET$ (GeV)    & $>250$ & $>300$ & $>350$ & $>400$ & $>500$ & $>600$ & $>700$ & $>800$ \\
$\sigma_\mathrm{vis}^{95}$ (fb) & 531  & 330   & 188   & 93   & 43   & 19    & 7.7  & 4.9  \\
\hline\hline
\end{tabular}
\end{table}

Moreover, various signal regions are defined according to further cuts on $\missET$, aiming at various models and different DM particle masses.
In each signal region, the ATLAS collaboration derives the observed upper limit at 95\% confidence level (C.L.) on the visible cross section $\sigma_\mathrm{vis}$, which is defined as the product of production cross section, acceptance, and efficiency.
Here we choose several inclusive signal regions tabulated in Table~\ref{tablesr}.
Based on simulation, we estimate the visible cross section in these signal regions for the inert sextuplet model and derive experimental bounds.

\begin{figure}[!t]
\centering
\includegraphics[width=0.49\textwidth]{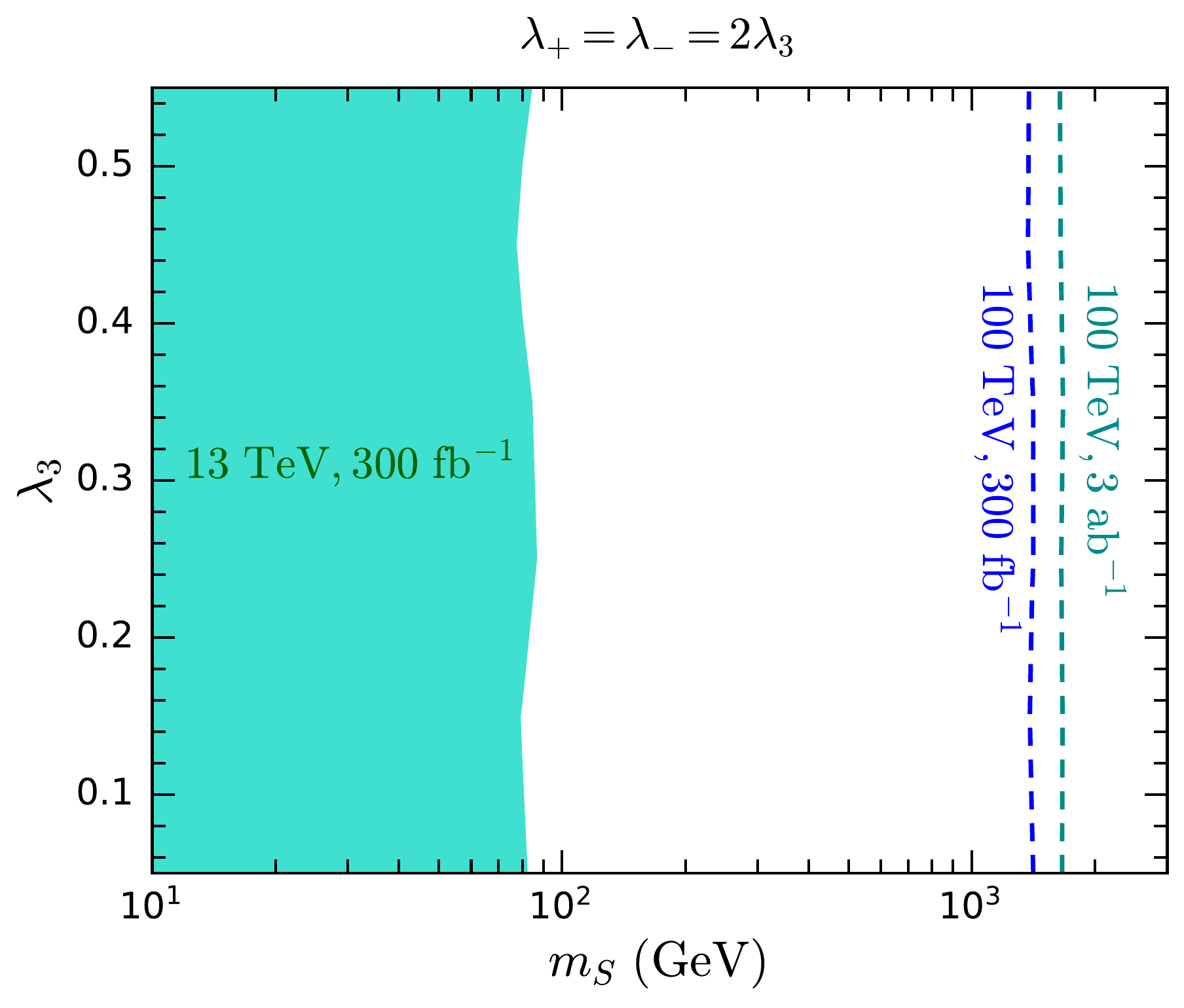}
\caption{Current constraint and future sensitivity of the $\text{monojet}+\missET$ channel in the $m_S$-$\lambda_3$ plane for $\lambda_+ = \lambda_- = 2\lambda_3$.
The turquoise region is excluded at 95\% C.L. by the 13~TeV ATLAS $\text{monojet}+\missET$ search with a dataset of $36.1~\si{fb^{-1}}$~\cite{Aaboud:2017phn}.
The blue and purple dashed lines denote the expected 95\% C.L. exclusion limits at a 100~TeV $pp$ collider with integrated luminosities $300~\si{fb^{-1}}$ and $3~\si{ab^{-1}}$, respectively.}
\label{fig:LHC:monojet:excluded}
\end{figure}

The 95\% C.L. exclusion region in the $m_S$-$\lambda_3$ plane for $\lambda_+ = \lambda_- = 2\lambda_3$ with the eight signal regions combined are shown in Fig.~\ref{fig:LHC:monojet:excluded}.
Such a result is obtained by analyzing the signal simulation samples for parameters in a $m_S$-$\lambda_3$ grid and performing appropriate interpolation.
Note that the parameter relation $\lambda_+ = \lambda_- = 2\lambda_3$ satisfies the condition~\eqref{eq:flat_cond}, and hence direct detection experiments give no constraint.
Furthermore, this relation leads to degenerate mass spectra at tree level with
\begin{equation}
m_{a^0} = m_{S^{\pm}_1} = m_{S^{2\pm}_1} = m_{S^{3\pm}} < m_{S^{2\pm}_2} = m_{S^{\pm}_2} = m_{\phi^0}.
\end{equation}
Although electroweak loop corrections break the degeneracy and slightly lift up the masses of $S^{\pm}_1$, $S^{2\pm}_1$, and $S^{3\pm}$~\cite{Cirelli:2005uq}, the mass splittings between these scalar bosons and $a^0$ are too small to induce hard visible particles that can be triggered in monojet searches.
Therefore, in this case these scalars also contribute to $\missET$, and one would expect a fairly well sensitivity in the monojet channel.
Nevertheless, Fig.~\ref{fig:LHC:monojet:excluded} shows that the 13~TeV monojet search only explore a region up to $m_S \sim 85~\si{GeV}$.
Although a small dependence on $\lambda_3$ manifests in the monojet constraint, it is not physical, but due to the statistical fluctuation in the signal simulation samples.

\subsubsection{100~TeV $pp$ collider sensitivity}

Since the $\text{monojet}+\missET$ search at the LHC does not explore the model well, we turn to study this channel at a 100~TeV $pp$ collider and estimate the improvement of the sensitivity.
We simulate signal and background event samples for $pp$ collisions at $\sqrt{s} = 100~\si{TeV}$.
Only two dominant SM backgrounds, $Z(\rightarrow\nu\bar{\nu})+\text{jets}$ and $W(\rightarrow \ell\nu)+\text{jets}$, are considered.
The 13~TeV ATLAS monojet analysis~\cite{Aaboud:2017phn} has shown that the events from these two backgrounds in the signal regions are more than the sum of the events from the other backgrounds by at least one order of magnitude.
It is reasonable to expect that the $Z(\rightarrow\nu\bar{\nu})+\text{jets}$ and $W(\rightarrow \ell\nu)+\text{jets}$ backgrounds would still be dominant at $\sqrt{s} = 100~\si{TeV}$.
Including the other backgrounds would only slightly reduce the sensitivity.

Then we require reconstructed jets to have $\pT > 80~\si{GeV}$ and $|\eta|<2.8$, and reconstructed electrons (muons) to have $\pT > 20~(10)~\si{GeV}$ and $|\eta|<2.47~(2.5)$.
Similar to the 13~TeV search, the following selection cuts are applied to the simulation samples.
\begin{itemize}
\item \textit{Cut~1.}---A leading jet with $\pT > 1.5~\si{TeV}$ and $|\eta|<2.4$.
\item \textit{Cut~2.}---No more than four reconstructed jets.
\item \textit{Cut~3.}---$\Delta\phi(j,\misspT) > 0.4$.
\item \textit{Cut~4.}---No reconstructed electron or muon.
\item \textit{Cut~5.}---$\missET > 1.6~\si{TeV}$.
\end{itemize}

\begin{table}[!t]
\renewcommand{\arraystretch}{1.2}
\setlength{\tabcolsep}{.5em}
\centering
\caption{Information of four BMPs.}
\label{tab:BMP}
\begin{tabular}{cccccc}
\hline
\hline
 & BMP1 & BMP2 & BMP3 & BMP4 \\ \hline
$\lambda_{+}$                                 &0.84           &0.41           &0.55          &0.345          \\ 
$\lambda_{-}$                                 &0.6            &0.35           &0.25          &0.075          \\ 
$\lambda_{3}$                                 &0.4            &0.2            &0.25          &0.15           \\ 
$m_S$ (GeV)                                  &250            &200            &300           &300            \\ 
$m_{a^0}$, $m_{\phi^0}$ (GeV)                       &250, 302.65     &200, 233.56     &300, 328.92    &300, 317.67     \\ 
$m_{S_{1}^{\pm}}$, $m_{S_{2}^{\pm}}$ (GeV)            &250.72, 302.06  &200.24, 233.35  &300.64, 328.33 &300.49, 317.21  \\ 
$m_{S_{1}^{2\pm}}$, $m_{S_{2}^{2\pm}}$ (GeV)          &252.96, 300.18  &200.97, 232.72  &302.77, 326.37 &302.14, 315.63  \\ 
$m_{S^{3\pm}}$ (GeV)  &257.17         &202.26         &307.48        &306.74         \\ 
$m_{\phi^0} - m_{a^0}$ (GeV)                          &$52.65$ &$33.56$ &$28.92$&$17.67$ \\ 
$\mathrm{BR}(\phi^{0}\rightarrow a^0 \ell^{+}\ell^{-})$    &3.37\%         &3.00\%               &4.64\%        &6.21\%         \\
$\mathrm{BR}(S^{\pm}_{2}\rightarrow S^{\pm}_{1}\ell^{+}\ell^{-})$ &2.99\%         &2.63\%               &4.16\%        &5.62\%         \\
$\mathrm{BR}(S^{2\pm}_{2}\rightarrow S^{2\pm}_{1}\ell^{+}\ell^{-})$&2.51\%         &2.17\%               &7.21\%        &7.23\%         \\
$S$                                           &0.0448         &0.0426         &0.0142        &0.00440        \\ 
$T$                                           &$-0.0222$        &$-0.00483$       &$-0.0113$       &$-0.00525$       \\ 
$U$                                           &$-0.00589$       &$-0.00208$       &$-0.00226$      &$-0.00108$       \\ 
\hline\hline
\end{tabular}
\end{table}

As illuminating examples, we consider four benchmark points (BMPs), whose detailed information is listed in Table~\ref{tab:BMP}.
All the four BMPs satisfy the condition~\eqref{eq:flat_cond}, leading to $m_{a^0} = m_S$, and there is no constraint from direct detection experiments.
The mass spectra in these BMPs have various degrees of compression, which can be represented by the largest mass splitting $m_{\phi^0} - m_{a^0}$.

In order to estimate the sensitivity at a future $pp$ collider, we define the signal significance as~\cite{Low:2014cba}
\begin{equation}
\mathcal{S}=\frac{N_\mathrm{S}}{\sqrt{N_\mathrm{B}+(\beta N_\mathrm{B})^{2}+(\gamma N_\mathrm{S})^{2}}},
\end{equation}
where $N_\mathrm{S}$ ($N_\mathrm{B}$) is the number of signal (total background) events.
$\beta$ and $\gamma$ represent the fractions of systematic uncertainties on $N_\mathrm{B}$ and on $N_\mathrm{S}$, respectively.
For the monojet channel at $\sqrt{s} = 100~\si{TeV}$, we assume $\beta = 1\%$ and $\gamma = 10\%$.

\begin{table}[!t]
\renewcommand{\arraystretch}{1.2}
\setlength{\tabcolsep}{.5em}
\centering
\caption{Visible cross section $\sigma_\mathrm{vis}$ in fb for the backgrounds and signal BMPs after each cut in the  $\text{monojet}+\missET$ channel at $\sqrt{s} = 100~\si{TeV}$. The signal significance $\mathcal{S}$ corresponds to integrated luminosity $3~\si{ab^{-1}}$ for $\beta = 1\%$ and $\gamma = 10\%$.}
\label{tablemono}
\begin{tabular}{cccccccccccccc}
\hline
\hline
         & $W\to\ell\nu$   &$Z\to\nu\bar{\nu}$ &\multicolumn{2}{c}{BMP1} & \multicolumn{2}{c}{BMP2} & \multicolumn{2}{c}{BMP3} & \multicolumn{2}{c}{BMP4} \\      
&$\sigma_{\mathrm{vis}}$&$\sigma_{\mathrm{vis}}$&$\sigma_{\mathrm{vis}}$&$\mathcal{S}$&$\sigma_{\mathrm{vis}}$&$\mathcal{S}$&$\sigma_{\mathrm{vis}}$&$\mathcal{S}$&$\sigma_{\mathrm{vis}}$&$\mathcal{S}$\\ \hline
Cut~1 &4620   &1135  &53.5 &0.926 &65.7 &1.13 &45.3 &0.784 &44.1 &0.763 \\
Cut~2 &3652   &922   &46.2 &1.00  &56.1 &1.22 &40.3 &0.878 &39.8 &0.867 \\
Cut~3 &1913   &537   &31.4 &1.27  &37.1 &1.50 &28.4 &1.15  &29.3 &1.19 \\
Cut~4 &1168   &537   &29.6 &1.71  &35.1 &2.02 &26.3 &1.52  &27.5 &1.59 \\
Cut~5 &48.7   &111   &16.8 &7.21  &19.7 &7.73 &16.1 &7.07  &17.1 &7.26 \\
\hline\hline
\end{tabular}
\end{table}

In Table~\ref{tablemono}, we list the visible cross sections for the backgrounds and the four signal BMPs after each cut applied, as well as the signal significances of the BMPs.
The veto on leptons in cut~4 reduces $\sim 39\%$ of the $W(\rightarrow \ell\nu)+\text{jets}$ background, since it often induces a hard lepton.
The signal significances of the four BMPs remarkably increase after applying cut~5, reaching above 7.
Overall, we find that these cuts efficiently suppress the backgrounds and increase $\mathcal{S}$.

Now we define four signal regions by separately requiring $\missET > 1.6,~ 1.8,~ 2,~ 2.5~\si{TeV}$.
The expected exclusion limit at 95\% C.L. combining these signal regions in the monojet channel with $\sqrt{s}=100~\si{TeV}$ has been shown in Fig.~\ref{fig:LHC:monojet:excluded} for $\lambda_+ = \lambda_- = 2\lambda_3$.
We find that the 100~TeV monojet search could explore the parameter space up to $m_S \sim 1.4~(1.65)~\si{TeV}$ for an integrated luminosity of $300~\si{fb^{-1}}$ ($3~\si{ab^{-1}}$).

\subsection{$\text{Soft-dilepton} + \text{jets} + \missET$ channel}

As discussed above, the mass spectrum of the inert sextuplet scalar model is typically compressed.
Consequently, leptons (electrons and muons) from off-shell $Z$ and $W$ bosons in the scalar decays are usually quite soft.
Nevertheless, it is still possible to make use of such soft leptons for searching for dark sector scalars~\cite{Giudice:2010wb,Gori:2013ala,Schwaller:2013baa,Han:2014kaa,Baer:2014kya}.
We will focus on a pair of same-flavor opposite-sign (SFOS) leptons (either $e^+e^-$ or $\mu^+\mu^-$) from off-shell $Z$ decays, because they are rather unique for signal-background discrimination.
The relevant scalar decay processes are $S_2^{2+}\to S_1^{2+}+ Z^*(\to \ell^+\ell^-)$, $S_2^{+}\to S_1^{+}+ Z^*(\to \ell^+\ell^-)$, and $\phi^0\to a^0 + Z^*(\to \ell^+\ell^-)$, whose Feynman diagrams are shown in Fig.~\ref{fig:decay:Z} and branching ratios are exemplified in Table~\ref{tab:BMP}.
Some additional jets are allowed for keeping more signal events.
In particular, one hard jet recoiling against the dark sector scalar pair is helpful for inducing a larger $\missET$ and making the soft leptons more easy to be triggered. 
Thus, the search channel we would like to study is $\text{soft-dilepton} + \text{jets} + \missET$.

The important SM backgrounds in this channel are $t\bar{t} + \text{jets}$, $tW + \text{jets}$, $VV + \text{jets}$ ($V=W^\pm,Z$), and $\tau^+\tau^- + \text{jets}$.
The top quark exclusively decays via the weak process $t\rightarrow b W$. If the two $W$ bosons from a $t\bar{t}$ pair decay leptonically, the final state would contain a pair of leptons associated with undetected neutrinos, mimicking the signal.
The $tW + \text{jets}$ background has a similar feature, but its production rate is lower.
Nonetheless, a veto on $b$-tagged jets would be able to efficiently suppress both the $t\bar{t} + \text{jets}$ and $tW + \text{jets}$ backgrounds.
Moreover, the $VV + \text{jets}$ and $\tau^+\tau^- + \text{jets}$ backgrounds can contribute to the $\text{soft-dilepton} + \text{jets} + \missET$ final state when the decays of the $W/Z$ bosons or the taus produce a SFOS lepton pair.

Utilizing the information from the SFOS lepton pair and $\misspT$, we can construct the $m_{\tau\tau}$ variable~\cite{Han:2014kaa,Baer:2014kya,Barr:2015eva,Aaboud:2017leg} for reducing the $\tau^+\tau^- + \text{jets}$ background.
When a tau pair recoils against an energetic jet, the taus are highly boosted and their decay products would be nearly parallel.
In this case, for leptonic tau decays, the momentum of a daughter neutrino is basically collinear to the momentum of the corresponding daughter lepton.
Thus, the 4-momentum of a neutrino $\nu_{i}$ can be expressed as $p_{\nu_i}^\mu = \xi_i p_{\ell_{i}}^\mu$, where $p_{\ell_{i}}^\mu$ is the 4-momentum of the related lepton $\ell_{i}$ and the lepton mass has been neglected.
Therefore, the total missing transverse momentum becomes $\slashed{\mathbf{p}}_\mathrm{T} = \xi_{1}\mathbf{p}_\mathrm{T}^{\ell_{1}}+\xi_{2}\mathbf{p}_\mathrm{T}^{\ell_{2}}$.
By solving the two equations provided by this relation, we obtain the values of $\xi_1$ and $\xi_2$ for given $\misspT$, $\mathbf{p}_\mathrm{T}^{\ell_{1}}$, and $\mathbf{p}_\mathrm{T}^{\ell_{2}}$.
Since the 4-momentum of each tau is $p_{\tau_i}^\mu = p_{\ell_{i}}^\mu + p_{\nu_i}^\mu = (1 + \xi_i) p_{\ell_{i}}^\mu$, the invariant mass squared of the $\tau^+\tau^-$ pair can be expressed as $m_{\tau\tau}^2 = 2(1+\xi_1)(1+\xi_2)\, p_{\ell_1}\cdot p_{\ell_2}$ after neglecting the tau mass.
The $m_{\tau\tau}$ variable is defined as $m_{\tau\tau} \equiv \operatorname{sign}(m_{\tau\tau}^2)\sqrt{|m_{\tau\tau}^2|}$.
By this definition, $m_{\tau\tau}$ can be either positive or negative in practice.
Nonetheless, $m_{\tau\tau}$ approximates the true invariant mass of a highly boosted $\tau^+\tau^-$ pair with leptonically decays.
We will see later that a veto on events with $0< m_{\tau\tau} < 200~\si{GeV}$ is helpful for suppressing both the $\tau^+\tau^- + \text{jets}$ and $VV + \text{jets}$ backgrounds at a 100~TeV $pp$ collider.

\subsubsection{LHC constraint}

The $\text{soft-dilepton} + \text{jets} + \missET$ channel has been used in the ATLAS search for electroweak production of supersymmetric particles with compressed mass spectra at $\sqrt{s} = 13~\si{TeV}$ with a dataset of $36.1~\si{fb^{-1}}$~\cite{Aaboud:2017leg}.
We make use of the related search results to constrain the inert sextuplet model.

In the ATLAS search, reconstructed electrons (muons) are required to have $\pT > 4.5~(4)~\si{GeV}$ and $|\eta| < 2.47~(2.5)$, while reconstructed jets should have $\pT > 20~\si{GeV}$ and $|\eta| < 4.5$.
The following selection conditions are further applied.
\begin{itemize}
\item A SFOS lepton pair ($e^+e^-$ or $\mu^+\mu^-$) and no other leptons.
The leading lepton $\ell_1$ has $\pT^{\ell_1} > 5~\si{GeV}$.
The subleading lepton $\ell_2$ has $\pT^{\ell_2} > 4.5~(4)~\si{GeV}$ if it is an electron (muon).
The angular distance $\Delta R\equiv\sqrt{\Delta\phi^2 + \Delta\eta^2}$ between the two leptons satisfies $\Delta R_{\ell\ell} \in (0.05,2)$.
The invariant mass of the two leptons satisfies $m_{\ell\ell} \in[1,3)\cup (3.2,60]~\si{GeV}$, where the $[3,3.2]~\si{GeV}$ window is excluded for evading contributions from $J/\psi$ decays.
\item At least one jet but no $b$-tagged jet. Each jet $j$ satisfies $\Delta\phi(j,\misspT) > 0.4$. The leading jet $j_1$ satisfies $\pT^{j_1} > 100~\si{GeV}$ and $\Delta\phi(j_1,\misspT)>2$.
\item $\missET > 200~\si{GeV}$, $m_\mathrm{T}^{\ell_1} < 70~\si{GeV}$, and $\missET/H_\mathrm{T}^\mathrm{lep} > \max(5,\,15-2m_{\ell\ell}/\si{GeV})$, where  $m_\mathrm{T}^{\ell_1} \equiv \sqrt{2(E_\mathrm{T}^{\ell_{1}}\missET-\mathbf{p}_\mathrm{T}^{\ell_{1}}\cdot\misspT)}$ and  $H_\mathrm{T}^\mathrm{lep} \equiv \pT^{\ell_1} + \pT^{\ell_2}$.
Veto on events with $m_{\tau\tau} \in [0, 160]~\si{GeV}$.
\end{itemize}
Then the ATLAS collaboration defines seven signal regions with different inclusive bins of $m_{\ell\ell}$, and obtains the 95\% C.L. observed upper limit on the visible cross section, as listed in Table~\ref{tablesoftsr}.

\begin{table}[!t]
\renewcommand{\arraystretch}{1.2}
\setlength{\tabcolsep}{.6em}
\centering
\caption{Signal regions in the ATLAS $\text{soft-dilepton} + \text{jets} + \missET$ analysis at $\sqrt{s} = 13~\si{TeV}$~\cite{Aaboud:2017leg}.
$\sigma_\mathrm{vis}^{95}$ denotes the observed 95\% C.L. upper limit on the visible cross section $\sigma_\mathrm{vis}$.}
\label{tablesoftsr}
\begin{tabular}{cccccccc}
\hline\hline 
$m_{\ell\ell}/\si{GeV}\in$       & $[1,3]$ & $[1,5]$ & $[1,10]$ & $[1,20]$ & $[1,30]$ & $[1,40]$ & $[1,60]$ \\
\hline 
$\sigma_\mathrm{vis}^{95}$ (fb) & 0.1     & 0.18    & 0.34     & 0.61     & 0.59     & 0.72     & 0.80 \\
\hline\hline
\end{tabular}
\end{table}

\begin{figure}[!t]
\centering
\includegraphics[width=0.49\textwidth]{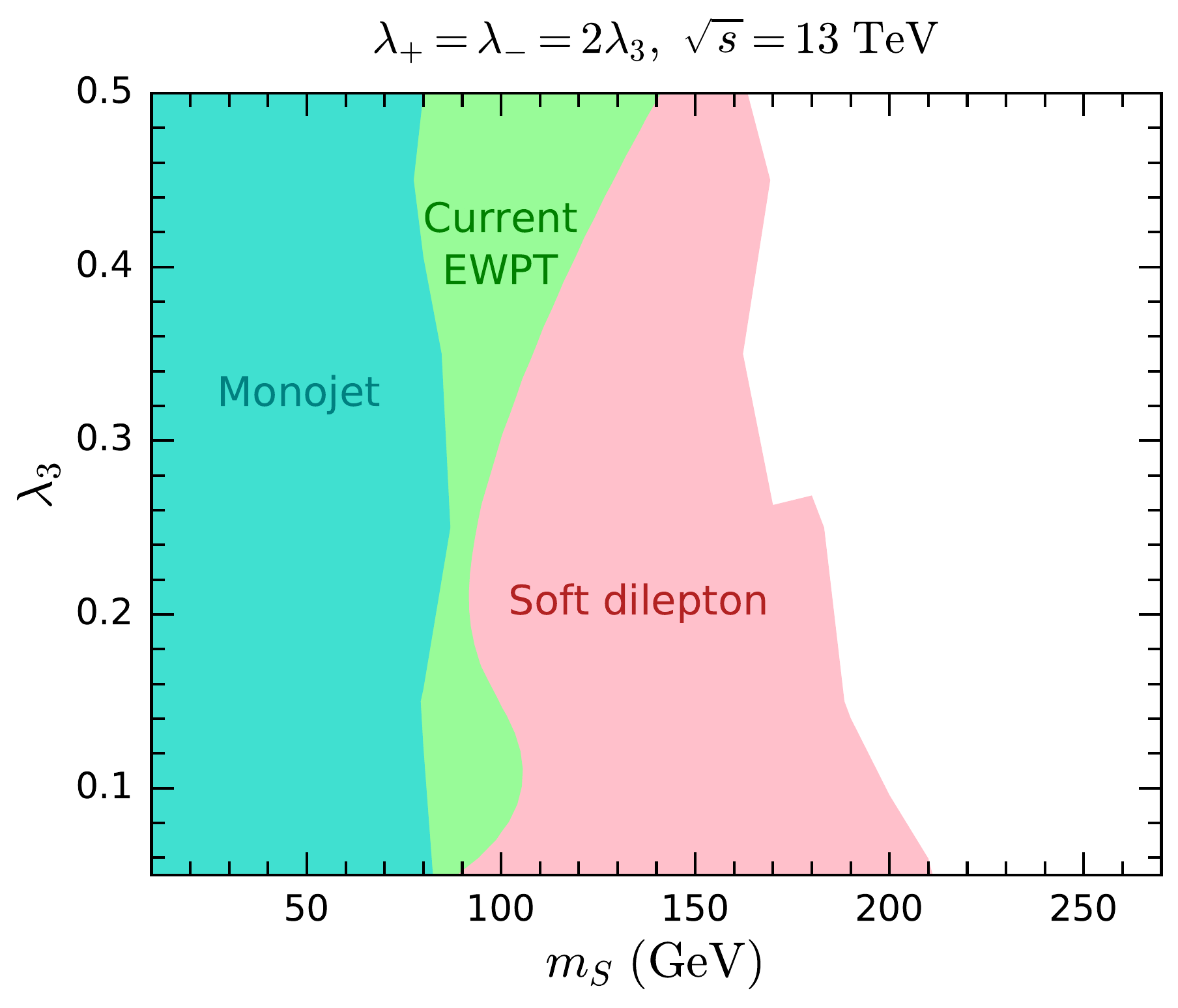}
\caption{Current constraints in the $m_S$-$\lambda_3$ plane for $\lambda_+ = \lambda_- = 2\lambda_3$.
The pink (turquoise) region is excluded at 95\% C.L. by the ATLAS $\text{soft-dilepton} + \text{jets} + \missET$ analysis~\cite{Aaboud:2017leg} ($\text{monojet}+\missET$ analysis~\cite{Aaboud:2017phn}) at $\sqrt{s} = 13~\si{TeV}$ with a dataset of $36.1~\si{fb^{-1}}$.
The light green region is excluded at 95\% C.L. by the global fit in the current EWPT from the Gfitter Group~\cite{Baak:2014ora} (see Sec.~\ref{sec:obliquepara}).}
\label{fig:LHC:soft:excluded}
\end{figure}

Combining these seven signal regions, we derive the 95\% C.L. exclusion region in the $m_S$-$\lambda_3$ plane for $\lambda_+ = \lambda_- = 2\lambda_3$, as presented in Fig.~\ref{fig:LHC:soft:excluded}.
We find that the $\text{soft-dilepton} + \text{jets} + \missET$ channel at the current LHC is more sensitive than the $\text{monojet}+\missET$ channel, excluding a region up to $m_S \sim 210~\si{GeV}$.
It seems that the soft-dilepton constraint becomes weak as $\lambda_3$ increases.
One reason is that larger $\lambda_3$ leads to larger mass splittings among the dark sector scalars, which result in harder leptons and hence larger $m_\mathrm{T}^{\ell_1}$ and $H_\mathrm{T}^\mathrm{lep}$.
Therefore, the selection conditions $m_\mathrm{T}^{\ell_1} < 70~\si{GeV}$ and $\missET/H_T^\mathrm{lep} > \max(5,\,15-2m_{\ell\ell}/\si{GeV})$ reject more signal events for larger $\lambda_3$ and weaken the constraint.
Besides, smaller mass splittings typically give rise to larger decay branching ratios into $\ell^+\ell^-$, which are helpful in increasing the signal significance for smaller $\lambda_3$.

\subsubsection{100~TeV $pp$ collider sensitivity}

Now we investigate the sensitivity of the  $\text{soft-dilepton} + \text{jets} + \missET$ channel at a 100~TeV $pp$ collider.
We require reconstructed leptons to have $\pT > 10~\si{GeV}$ and $|\eta| < 2.5$, and reconstructed jets to have $\pT > 40~\si{GeV}$ and $|\eta| < 4.5$.
We apply the following selection cuts to the simulation samples.
\begin{itemize}
\item \textit{Cut~1.}---A SFOS lepton pair and no other leptons; $\missET > 300~\si{GeV}$; at least one jet;  $\Delta\phi(j,\misspT) > 0.4$ for any jet $j$; $\pT^{j_1} > 240~\si{GeV}$ and $\Delta\phi(j_1,\misspT)>2$ for the leading jet $j_1$.
\item \textit{Cut~2.}---$0.05<\Delta R_{\ell\ell}<2$.
\item \textit{Cut~3.}---$m_{\tau\tau}<0$ or $m_{\tau\tau} > 200~\si{GeV}$.
\item \textit{Cut~4.}---No $b$-tagged jet.
\item \textit{Cut~5.}---$m_{\ell\ell} \in[1,3)\cup (3.2,70]~\si{GeV}$.
\end{itemize}
We find that the $m_\mathrm{T}^{\ell_1}$ and $\missET/H_T^\mathrm{lep}$ cuts used in the 13~TeV ATLAS  analysis are not helpful for the inert sextuplet model,
so these two cuts are abandoned for $\sqrt{s} = 100~\si{TeV}$.

\begin{figure}[!t]
\centering
\subfigure[~$\Delta R_{\ell\ell}$ distributions\label{distributionsoft:a}]
{\includegraphics[width=0.49\textwidth]{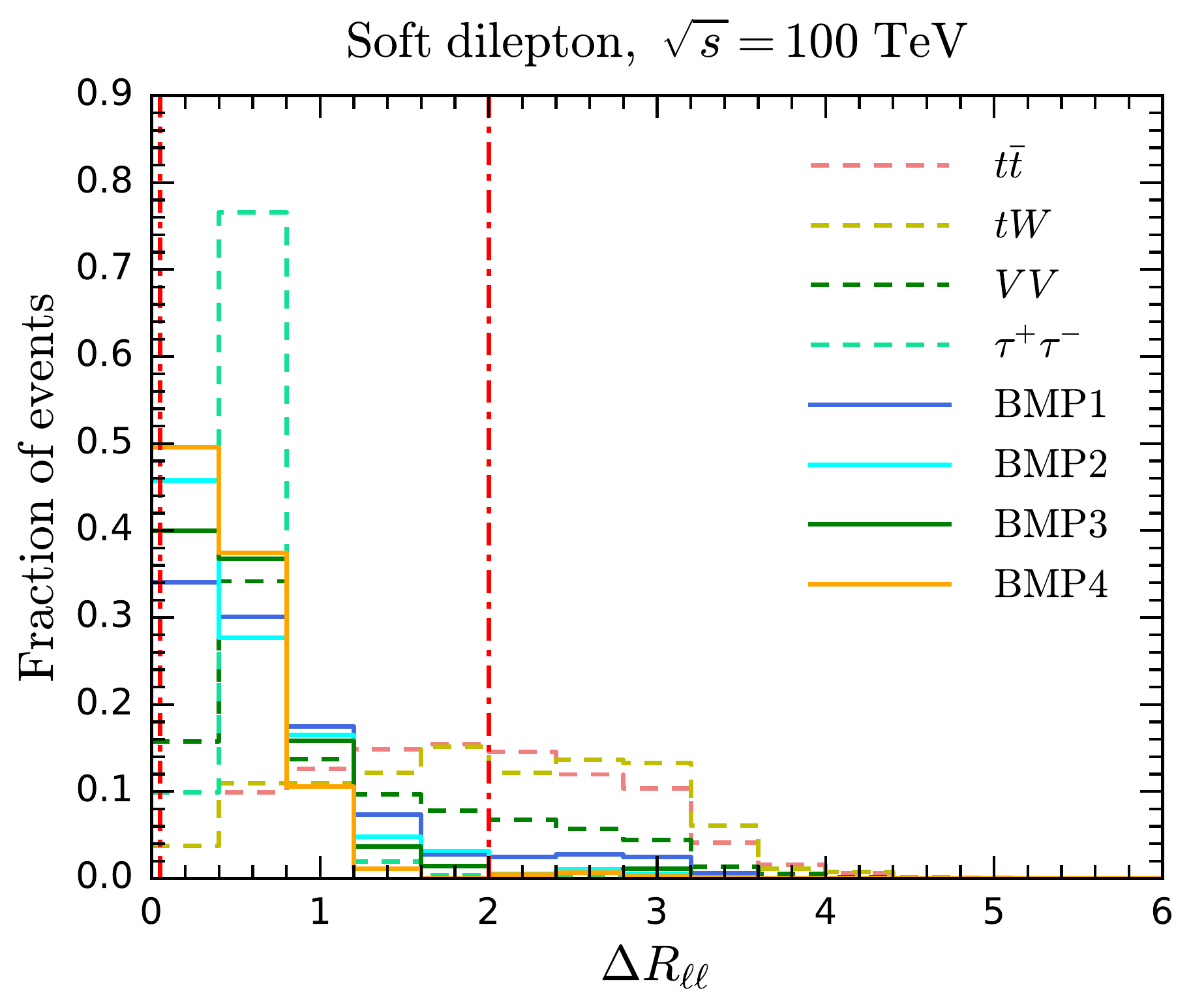}}
\subfigure[~$m_{\tau\tau}$ distributions\label{distributionsoft:b}]
{\includegraphics[width=0.49\textwidth]{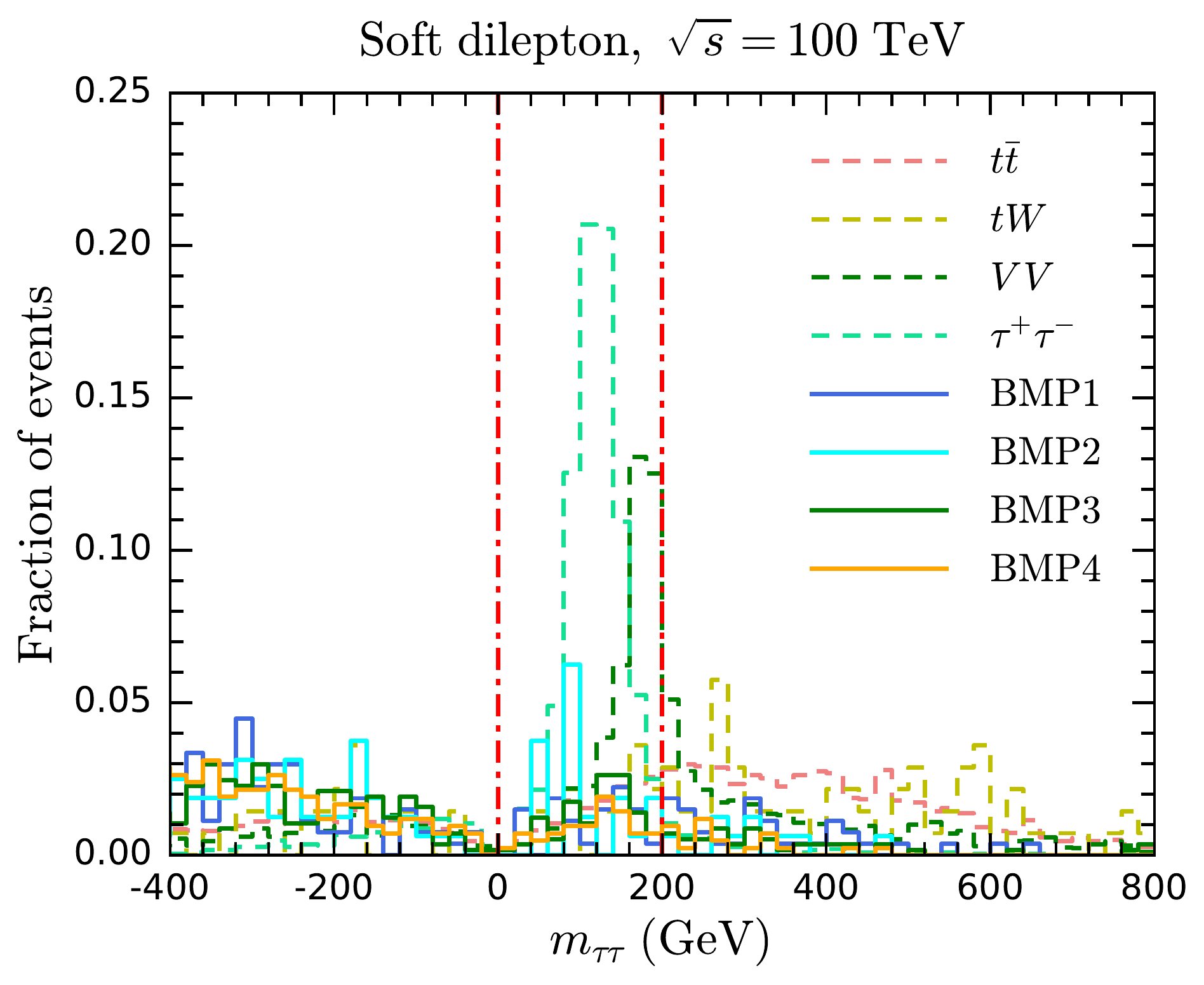}}
\subfigure[~$b$-jet number distributions\label{distributionsoft:c}]
{\includegraphics[width=0.49\textwidth]{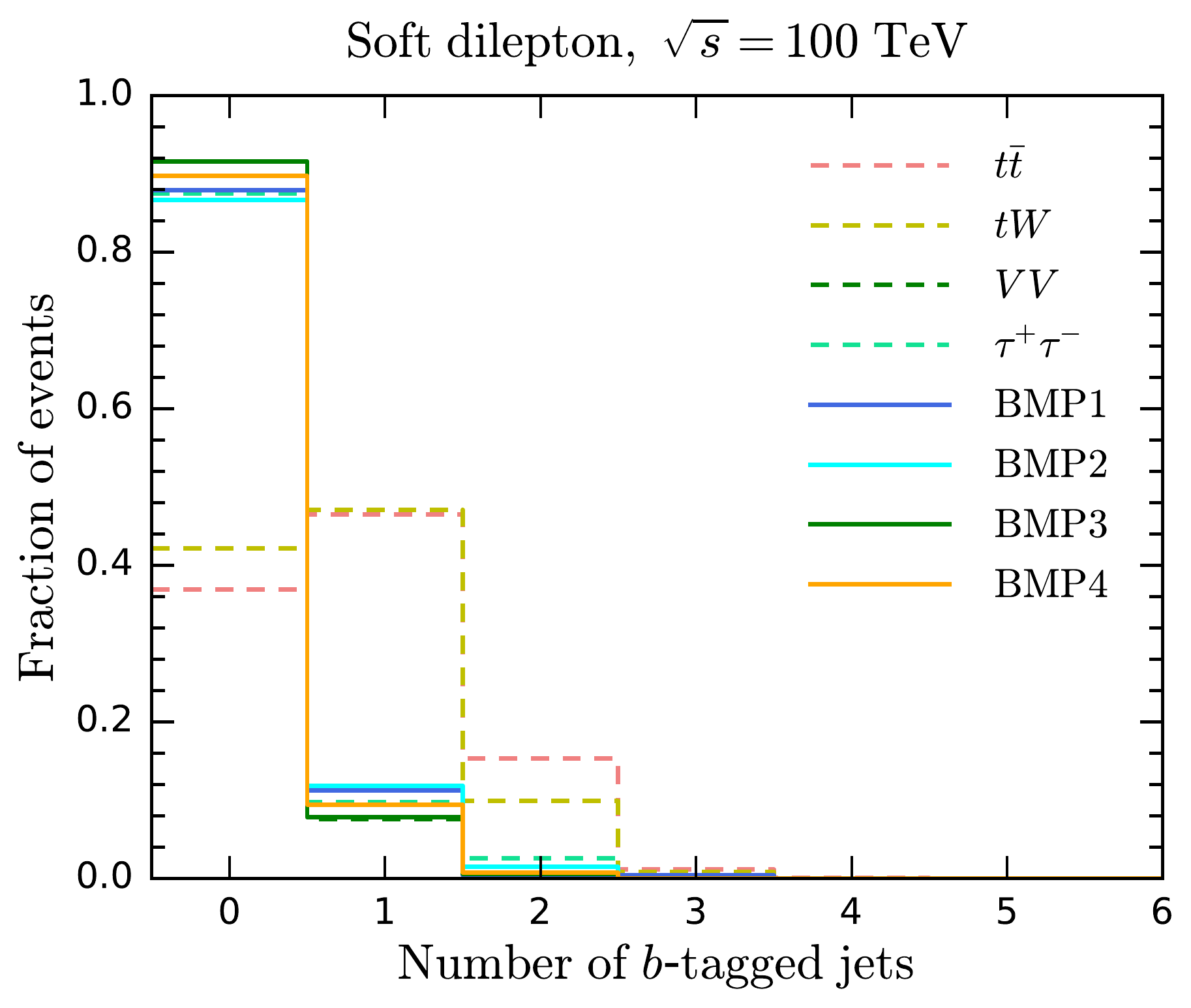}}
\subfigure[~$m_{\ell\ell}$ distributions\label{distributionsoft:d}]
{\includegraphics[width=0.49\textwidth]{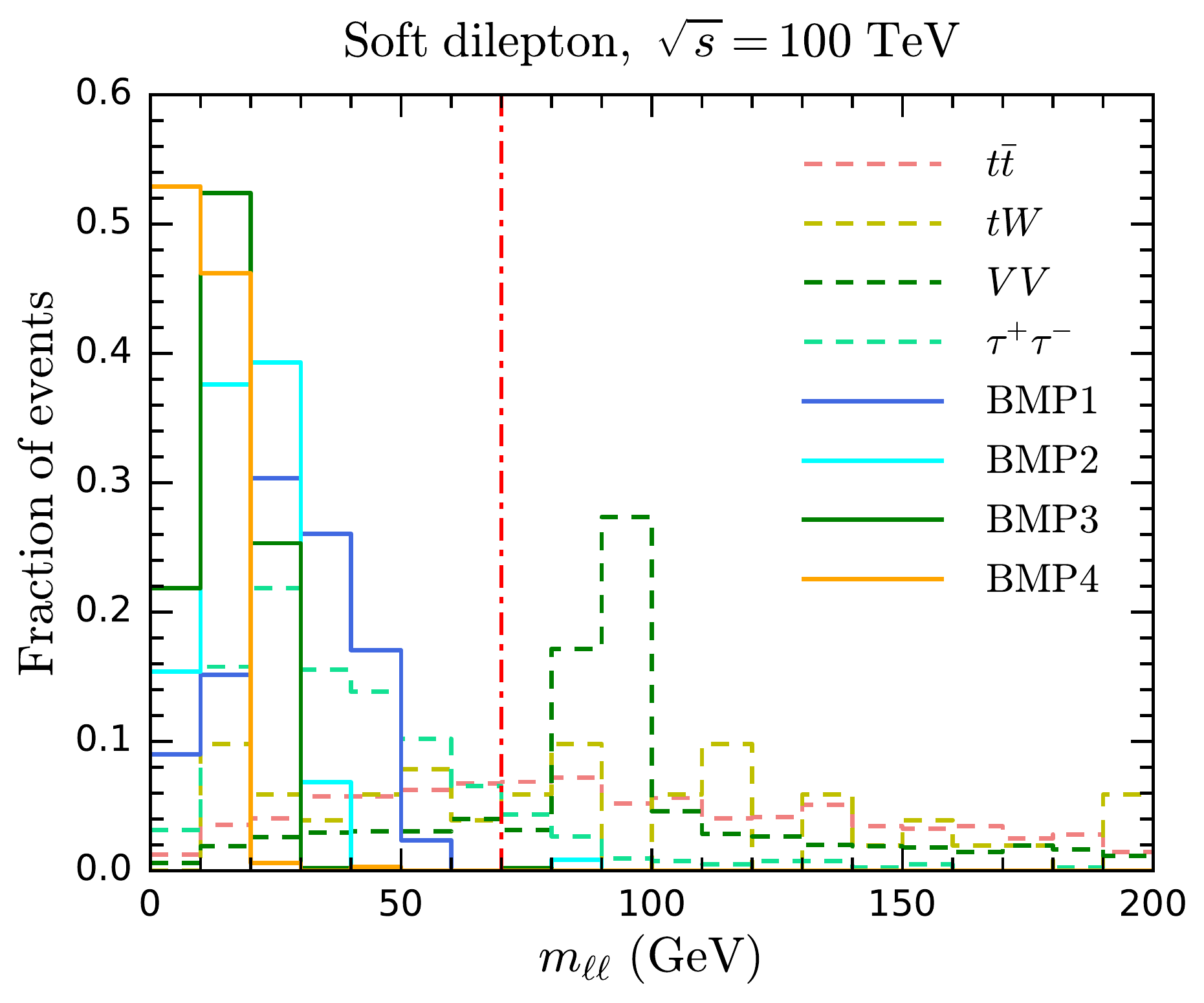}}
\caption{Fractions of background and signal events binned in $\Delta R_{\ell\ell}$ after cut~1 (a), in $m_{\tau\tau}$ after cut~2 (b), in the number of $b$-tagged jets after cut~3 (c), and in $m_{\ell\ell}$ after cut~4 (d) for the $\text{soft-dilepton} + \text{jets} + \missET$ channel at $\sqrt{s} = 100~\si{TeV}$.
The red dot-dashed lines indicate the cut thresholds.}
\label{distributionsoft}
\end{figure}

In order to justify these cut conditions, we demonstrate the distributions of backgrounds and signal BMPs in Fig.~\ref{distributionsoft}.
These distributions are expressed as the fractions of events binned in kinematic variables.
Moreover, the visible cross sections and the BMP signal significances are tabulated in Table~\ref{tab:cut:soft}.
For the $\text{soft-dilepton} + \text{jets} + \missET$ channel at $\sqrt{s} = 100~\si{TeV}$, we assume the fractions of the systematic uncertainties to be $\beta = 1\%$ and $\gamma = 10\%$.

\begin{table}[!t]
\renewcommand{\arraystretch}{1.2}
\setlength{\tabcolsep}{.3em}
\centering
\caption{Visible cross section $\sigma_\mathrm{vis}$ in fb for the backgrounds and signal BMPs after the cuts and for the signal regions in the  $\text{soft-dilepton} + \text{jets} + \missET$ channel at $\sqrt{s} = 100~\si{TeV}$. The signal significance $\mathcal{S}$ corresponds to integrated luminosity $3~\si{ab^{-1}}$ for $\beta = 1\%$ and $\gamma = 10\%$.}
\label{tab:cut:soft}
\begin{tabular}{ccccccccccccc}
\hline
\hline
     &$t\bar{t}$   &$tW$    &$VV$  &$\tau^{+}\tau^{-}$ & \multicolumn{2}{c}{BMP1} & \multicolumn{2}{c}{BMP2} & \multicolumn{2}{c}{BMP3} & \multicolumn{2}{c}{BMP4} \\
     &$\sigma_{\mathrm{vis}}$&$\sigma_{\mathrm{vis}}$&$\sigma_{\mathrm{vis}}$&$\sigma_{\mathrm{vis}}$&$\sigma_{\mathrm{vis}}$&$\mathcal{S}$&$\sigma_{\mathrm{vis}}$&$\mathcal{S}$&$\sigma_{\mathrm{vis}}$&$\mathcal{S}$&$\sigma_{\mathrm{vis}}$&$\mathcal{S}$\\ \hline
Cut~1    &2026 &354  &374  &111      &7.97 &0.278   &8.78 &0.306   &10.1 &0.352   &7.75 &0.270\\
Cut~2    &1140 &186  &303  &111      &6.55 &0.376   &7.47 &0.429   &9.30 &0.534   &7.13 &0.409 \\
Cut~3    &970  &162  &175  &20.7     &5.87 &0.441   &6.30 &0.474   &8.15 &0.611   &6.50 &0.488 \\
Cut~4    &358  &68.3 &160  &18.1     &5.16 &0.847   &5.46 &0.897   &7.46 &1.22    &5.84 &0.957 \\
Cut~5    &119  &25.5 &29.1 &15.7     &5.16 &2.61    &5.37 &2.71    &7.41 &3.62    &5.82 &2.91\\ \hline
SR-60 &94.8 &22.8 &22.6 &14.6     &5.16 &3.13    &5.37 &3.25    &7.41 &4.28    &5.82 &3.49 \\
SR-40 &51.9 &13.4 &12.9 &10.2     &4.15 &\textbf{4.19}    &5.37 &5.12    &7.41 &6.35    &5.80 &5.41 \\
SR-20 &16.8 &6.70 &3.98 &3.43     &1.25 &3.58    &2.85 &\textbf{6.59}    &5.51 &\textbf{8.61}    &5.77 &\textbf{8.71}\\
\hline\hline
\end{tabular}
\end{table}

Firstly, Fig.~\ref{distributionsoft:a} shows the $\Delta R_{\ell\ell}$ distributions after applying cut~1.
As we can see, the $t\bar{t}+ \text{jets}$ and $tW+ \text{jets}$ backgrounds tend to have much larger $R_{\ell\ell}$ than the BMPs and the other backgrounds.
A part of the $VV+ \text{jets}$ background exhibits a similar behavior.
This is because the two leptons come from decays of different particles in $t\bar{t}$, $tW$, and $W^+W^-$ and prefer to fly in opposite directions.
For the BMPs and the $VV+ \text{jets}$ and $\tau^+\tau^- + \text{jets}$ backgrounds, the SFOS lepton pair could come from an on- or off-shell $Z$ boson, which is sufficiently boosted due to the requirements of $\missET > 300~\si{GeV}$ and $p_\mathrm{T}^{j_1} > 240~\si{GeV}$ in cut~1, and give small $\Delta R_{\ell\ell}$.
Thus, the $\Delta R_{\ell\ell}$ cut is rather useful for suppressing the $t\bar{t} + \text{jets}$ and $tW + \text{jets}$ backgrounds, and also reduces the $VV+ \text{jets}$ background.
From Table~\ref{tab:cut:soft}, we find that the $t\bar{t} + \text{jets}$, $tW + \text{jets}$, and $VV+ \text{jets}$ backgrounds lose roughly $44\%$, $47\%$, and $19\%$ of events after we require $0.05<\Delta R_{\ell\ell}<2$ in cut~2, respectively.

Secondly, in Fig.~\ref{distributionsoft:b}, we present the $m_{\tau\tau}$ distributions after cut~2.
As discussed above, $m_{\tau\tau}$ approximates the invariant mass of the tau pair in the $\tau^+\tau^- + \text{jets}$, resulting in a peak at $m_{\tau\tau} \sim m_Z$.
In addition, the tau pair from the $W^+W^- \to \tau^+\tau^-\nu_\tau\bar\nu_\tau$ process in the $VV+ \text{jets}$ background leads to a peak around $\sim 2m_W$.
Therefore, the veto on $0\leq m_{\tau\tau}\leq 200~\si{GeV}$ in cut~3 significantly reduces the  $\tau^+\tau^- + \text{jets}$ and $VV+ \text{jets}$ backgrounds.
As shown in Table~\ref{tab:cut:soft}, only $\sim 19\%$ and $\sim 58\%$ of events in the $\tau^+\tau^- + \text{jets}$ and $VV+ \text{jets}$ backgrounds remain after cut~3, respectively.

Thirdly, the distributions of the number of $b$-tagged jets after cut~3 are illustrated in Fig.~\ref{distributionsoft:c}.
Since the jets induced by the $b$ quarks from the $t\bar{t} + \text{jets}$ and $tW + \text{jets}$ backgrounds have a high probability to be tagged as $b$-jets, most events from these two backgrounds have at least one $b$-tagged jet.
As a result, the veto on $b$-tagged jets in cut~4 removes $\sim 63\%$ ($\sim 58\%$) of event in the $t\bar{t} + \text{jets}$ ($tW + \text{jets}$) background.

Finally, we demonstrate the $m_{\ell\ell}$ distributions in Fig.~\ref{distributionsoft:d}.
As expected, the $VV+ \text{jets}$ background peaks around $m_{\ell\ell} \sim m_Z$, manifesting the $Z$ pole.
Furthermore, the $t\bar{t} + \text{jets}$ and $tW + \text{jets}$ backgrounds tend to have larger $m_{\ell\ell}$ than the BMPs.
Consequently, the requirement of $m_{\ell\ell} \in[1,3)\cup (3.2,70]~\si{GeV}$ in cut~5 kills about $82\%$, $67\%$, and $63\%$ of events in the $VV+ \text{jets}$, $t\bar{t} + \text{jets}$, and $tW + \text{jets}$ backgrounds, respectively.
On the other hand, since the SFOS lepton pair in the signal BMPs comes from the decay processes $\phi^{0}\rightarrow a^0 \ell^{+}\ell^{-}$, $S^{\pm}_{2}\rightarrow S^{\pm}_{1}\ell^{+}\ell^{-}$, and $S^{2\pm}_{2}\rightarrow S^{2\pm}_{1}\ell^{+}\ell^{-}$, its invariant mass $m_{\ell\ell}$ reflects the mass splittings of the dark sector scalars.
As shown in Table~\ref{tab:BMP}, BMP1, BMP2, BMP3, and BMP4 have descending $m_{\phi^0} - m_{a^0}$.
Therefore, descending peaks exhibit accordingly in their $m_{\ell\ell}$ distributions.

Here we define six signal regions by requiring $m_{\ell\ell} \in [1,10]$, $[1,20]$, $[1,40]$, $[1,50]$, $[1,60]$, and $[1,70]$, dubbed SR-10, SR-20, SR-40, SR-50, SR-60, and SR-70, respectively.
Table~\ref{tab:cut:soft} also lists the visible cross sections and the signal significances of the four BMPs in the signal regions SR-60, SR-40, and SR-20.
We find that BMP1 has the largest $\mathcal{S}$ in SR-40, while the other three BMPs reach the largest $\mathcal{S}$ in SR-20.

\begin{figure}[!t]
\centering
\includegraphics[width=0.49\textwidth]{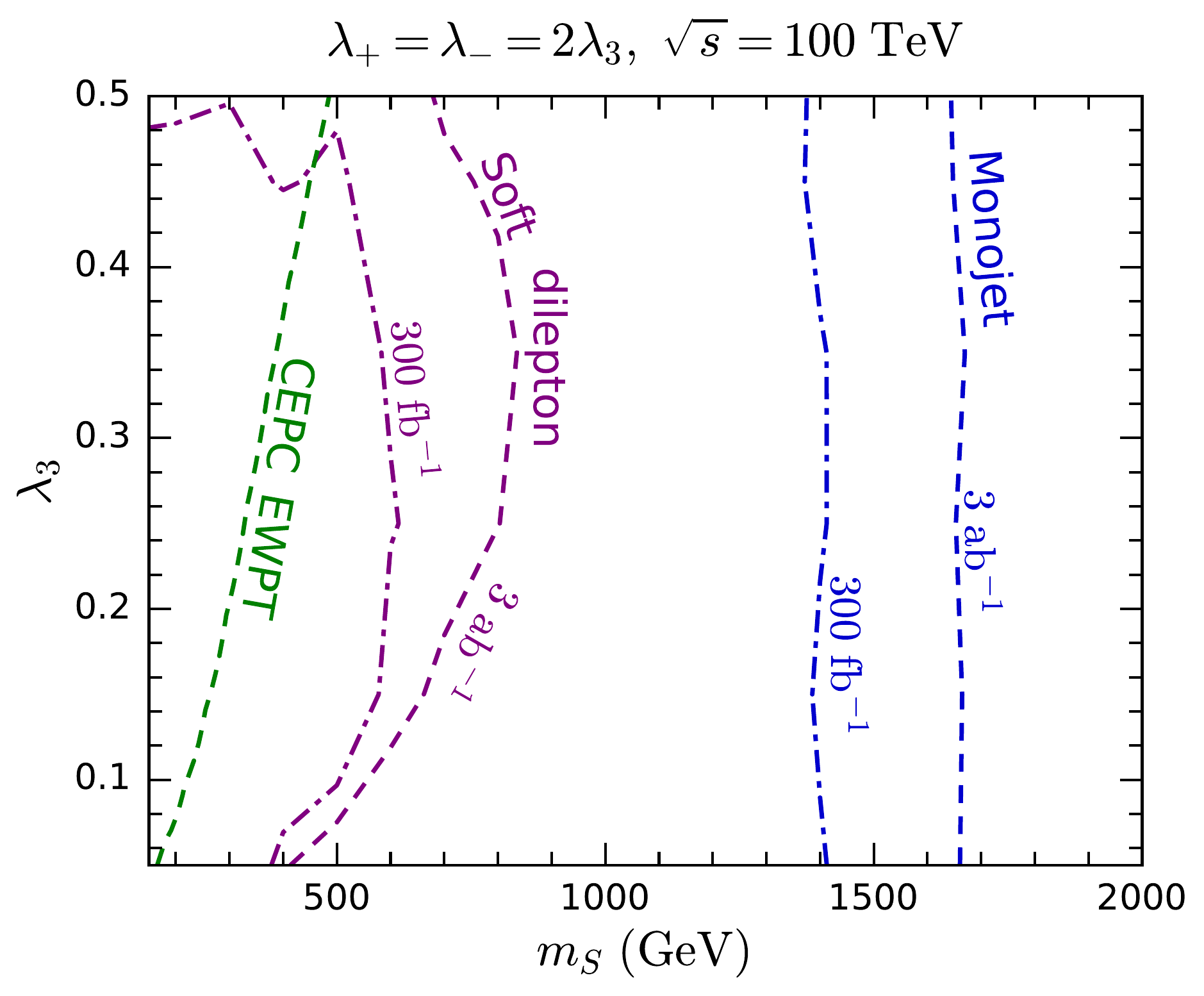}
\caption{Sensitivities of future $pp$ and $e^+e^-$ colliders presented in the $m_S$-$\lambda_3$ plane for $\lambda_+ = \lambda_- = 2\lambda_3$.
The purple (blue) lines denote the expected 95\% C.L. exclusion limits of the $\text{soft-dilepton} + \text{jets} + \missET$ ($\text{monojet}+\missET$) channel at a 100~TeV $pp$ collider with integrated luminosities $300~\si{fb^{-1}}$ and $3~\si{ab^{-1}}$.
The green line indicates the expected 95\% C.L. exclusion limit of the future EWPT at the CEPC~\cite{CEPCStudyGroup:2018ghi} (see Sec.~\ref{sec:obliquepara}).}
\label{fig:future:soft}
\end{figure}

In Fig.~\ref{fig:future:soft}, we show the expected 95\% C.L. exclusion limit in the $\text{soft-dilepton} + \text{jets} + \missET$ channel with the six signal regions combined at $\sqrt{s} = 100~\si{TeV}$, presented in the $m_S$-$\lambda_3$ plane for $\lambda_+ = \lambda_- = 2\lambda_3$.
With an integrated luminosity of $300~\si{fb^{-1}}$ ($3~\si{ab^{-1}}$), the 100~TeV soft-dilepton search could probe a region up to $m_S \sim 600~(840)~\si{GeV}$.
Although the soft-dilepton channel is more powerful than the monojet channel at the 13~TeV LHC, it becomes less sensitive at a 100~TeV machine.
The reason is that larger $m_S$ corresponds to smaller mass splittings among the dark sector scalars, as demonstrated in Figs.~\ref{fig:masssplit} and \ref{fig:split:lam3}.
Therefore, for $m_S \sim \si{TeV}$, the leptons from the scalar decays are too soft to pass the selection cuts, resulting in the loss of sensitivity in the soft-dilepton channel.
On the other hand, the monojet search becomes more sensitive for $m_S \sim \si{TeV}$, because it requires no reconstructed leptons in the final state.

\section{Indirect probe with electroweak oblique parameters}
\label{sec:obliquepara}

Since the sextuplet components participate electroweak gauge interactions, they could contribute to several electroweak precision observables at one-loop level~\cite{Earl:2013fpa}.
Most of the effects come from vacuum polarization diagrams of the electroweak gauge bosons and can be incorporated into the electroweak oblique parameters $S$, $T$, and $U$~\cite{Peskin:1990zt,Peskin:1991sw}, whose values can be determined in the EWPT.
In this section, we explore the current EWPT constraint on the inert sextuplet model, as well as the future sensitivity.

$S$, $T$, and $U$ are linear combinations of the $g^{\mu\nu}$ coefficients $\Pi_{PQ}(p^2)$ of the gauge boson vacuum polarizations contributed by new physics, defined as
\begin{eqnarray}
S &=& \frac{4s_\mathrm{W}^{2}c_\mathrm{W}^{2}}{\alpha}\left[\Pi_{ZZ}'(0)-\frac{c_\mathrm{W}^{2}-s_\mathrm{W}^{2}}{s_\mathrm{W}c_\mathrm{W}}\Pi_{ZA}'(0)-\Pi_{AA}'(0)\right], \\
T &=& \frac{1}{\alpha}\left[\frac{\Pi_{WW}(0)}{m_{W}^{2}}-\frac{\Pi_{ZZ}(0)}{m_{Z}^{2}}\right], \\
U &=& \frac{4s_\mathrm{W}^{2}}{\alpha}\big[\Pi_{WW}'(0)-c_\mathrm{W}^{2}\Pi_{ZZ}'(0)-2s_\mathrm{W}c_\mathrm{W}\Pi_{ZA}'(0)-s_\mathrm{W}^{2}\Pi_{AA}'(0)\big],
\end{eqnarray}
where $\alpha$ is the fine structure constant, $s_\mathrm{W} \equiv \sin\theta_\mathrm{W}$, and $\Pi'_{PQ}(0)\equiv\partial\Pi_{PQ}(p^{2})/\partial p^{2}|_{p^{2}=0}$.
Note that the SM predicts $S = T = U = 0$.
The one-loop contributions to $\Pi_{PQ}(p^2)$ by the dark sector bosons in the inert sextuplet model are given in Appendix~\ref{app:Pi}.

A $\SUtwoL$ multiplet with nonzero hypercharge split by the Higgs VEV typically gives nonzero $S$, $T$, and $U$~\cite{Zhang:2006de,Zhang:2006vt}.
Nevertheless, if the interactions between the multiplet and the Higgs doublet respect a global $\SUtwoL\times\SUtwoR$ symmetry, 
a custodial $\SUtwo_\mathrm{L+R}$ symmetry~\cite{Sikivie:1980hm} remains after electroweak symmetry breaking, resulting in vanishing $T$ and $U$.
In the inert sextuplet scalar model, the custodial symmetry corresponds to the condition
\begin{equation}
\lambda_- = \pm 2\lambda_3.
\end{equation}

To make this clear, we construct two $\SUtwoR$ doublet $\mathbf{H}^{I}$ and $\mathbf{S}^{I}$ ($I = 1,2$), whose components are $\SUtwoL$ doublets and sextuplets, respectively, given by
\begin{eqnarray}
\mathbf{H}^{1,i} &=& \epsilon^{ij}H_j^\dag,\quad
\mathbf{H}^{2,i}=H^{i}, \\
\mathbf{S}^{1,ijklm} &=& \epsilon^{ip}\epsilon^{jq}\epsilon^{kr}\epsilon^{ls}\epsilon^{mt}S_{pqrst}^\dag,\quad
\mathbf{S}^{2,ijklm}=S^{ijklm}.
\label{eq:S_bf}
\end{eqnarray}
Thus, a generic potential respecting the global $\SUtwoL\times\SUtwoR$ symmetry can be written down as
\begin{eqnarray}
V_{\mathrm{cus}} &=& \lambda_{a}\,\mathbf{H}_{I,i}^{\dag}\mathbf{H}^{I,i}\mathbf{S}_{J,jklmn}^{\dag}\mathbf{S}^{J,jklmn} 
+\lambda_{b}\,\mathbf{H}_{I,i}^{\dag}\mathbf{S}_{J,pjklm}^{\dag}\mathbf{S}^{J,jklmn}\mathbf{H}^{I,q}\epsilon^{ip}\epsilon_{nq}
\nonumber\\
&& +\lambda_{c}\,\mathbf{H}_{I,i}^{\dag}\mathbf{S}^{I,ijklm}\mathbf{S}_{J,jklmn}^{\dag}\mathbf{H}^{J,n}
+\lambda_{d}\,\mathbf{H}_{I,i}^{\dag}\mathbf{S}^{J,ijklm}\mathbf{S}_{K,jklmn}^{\dag}\mathbf{H}^{L,n}\epsilon^{IK}\epsilon_{JL}
\nonumber\\
&=& 4\lambda_{a}H_i^\dag {H^i}S_{jklmn}^\dag {S^{jklmn}}
+2(\lambda_b + \lambda_c)H^{\dag}_{i} S^{iklmn} S^{\dag}_{jklmn} H^{j}
\nonumber\\
&&
+2(\lambda_b + \lambda_d)H^{\dag}_{i} S^{\dag}_{kmnpq} S^{lmnpq} H^{j}\epsilon^{ik}\epsilon_{lj}
\nonumber\\
&& +(\lambda_c - \lambda_d)(H^{\dag}_{i}H^{\dag}_{j}S^{iklmn}S^{jpqrs}\epsilon_{kp}\epsilon_{lq}\epsilon_{mr}\epsilon_{ns}+\mathrm{H.c.}). 
\end{eqnarray}
After electroweak symmetry breaking, such a potential would respect the custodial $\SUtwo_\mathrm{L+R}$ symmetry.
As discussed above, the $4\lambda_{a}H_i^\dag {H^i}S_{jklmn}^\dag {S^{jklmn}}$ term is not independent.
Compared with the potential \eqref{eq:potential}, we find that
\begin{equation}
{\lambda _1} = 2({\lambda _b} + {\lambda _c}),\quad
{\lambda _2} = 2({\lambda _b} + {\lambda _d}),\quad
{\lambda _3} = {\lambda _c} - {\lambda _d},
\end{equation}
which lead to ${\lambda _ - } = {\lambda _1} - {\lambda _2} = 2{\lambda _3}$.
Therefore, $\lambda_- = 2\lambda_3$ is a condition for respecting the custodial symmetry.
On the other hand, if we define
\begin{equation}
\mathbf{S}^{1,ijklm} = -\epsilon^{ip}\epsilon^{jq}\epsilon^{kr}\epsilon^{ls}\epsilon^{mt}S_{pqrst}^\dag,\quad
\mathbf{S}^{2,ijklm}=S^{ijklm},
\end{equation}
instead of Eq.~\eqref{eq:S_bf}, we would prove that $\lambda_- = -2\lambda_3$ is another condition for the custodial symmetry.

\begin{figure}[!t]
\centering
\subfigure[~$\lambda_+ = \lambda_3 = 0.5$\label{fig:STU:lam-}]
{\includegraphics[width=0.49\textwidth]{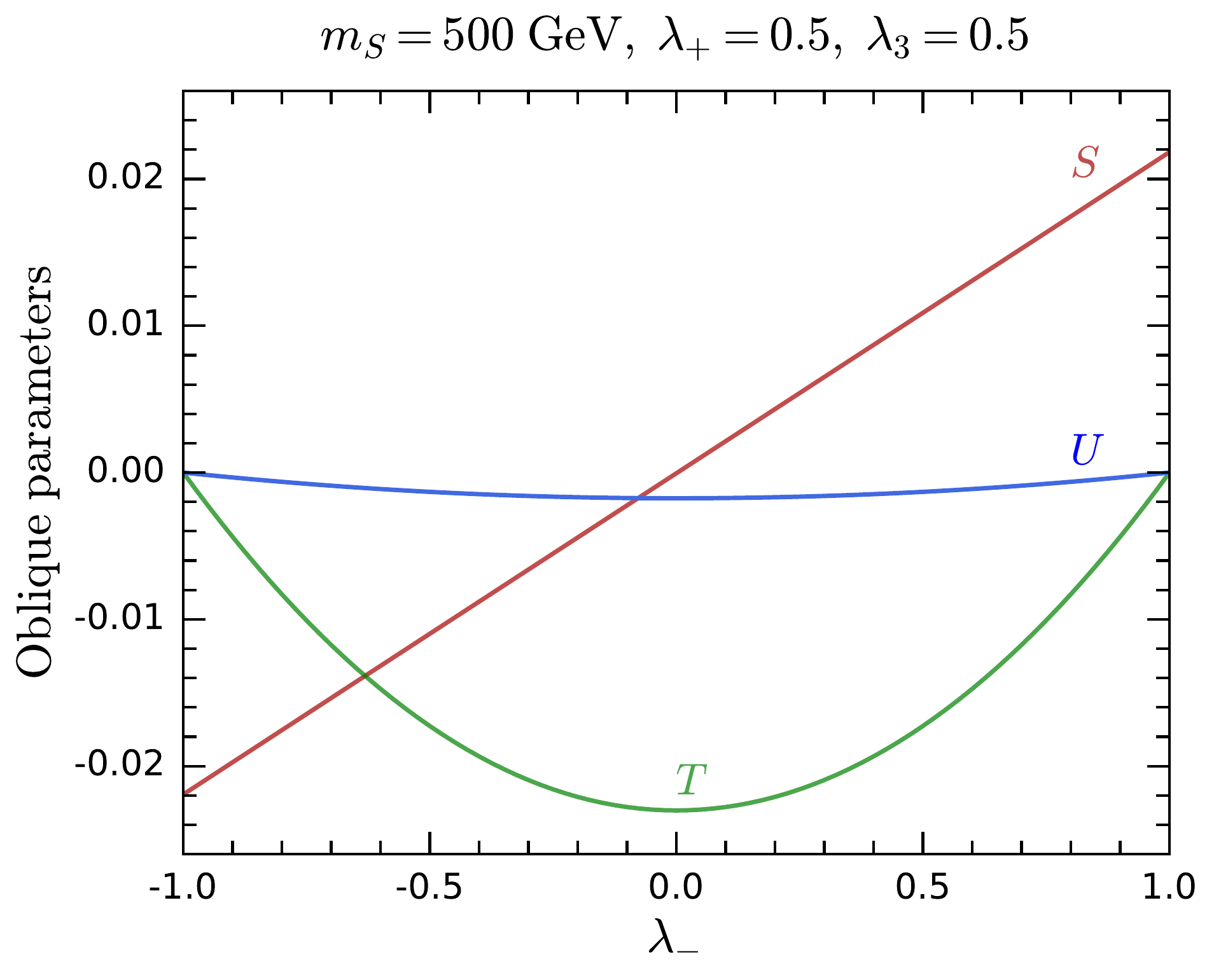}}
\subfigure[~$\lambda_+ = \lambda_- = 0.5$\label{fig:STU:lam3}]
{\includegraphics[width=0.49\textwidth]{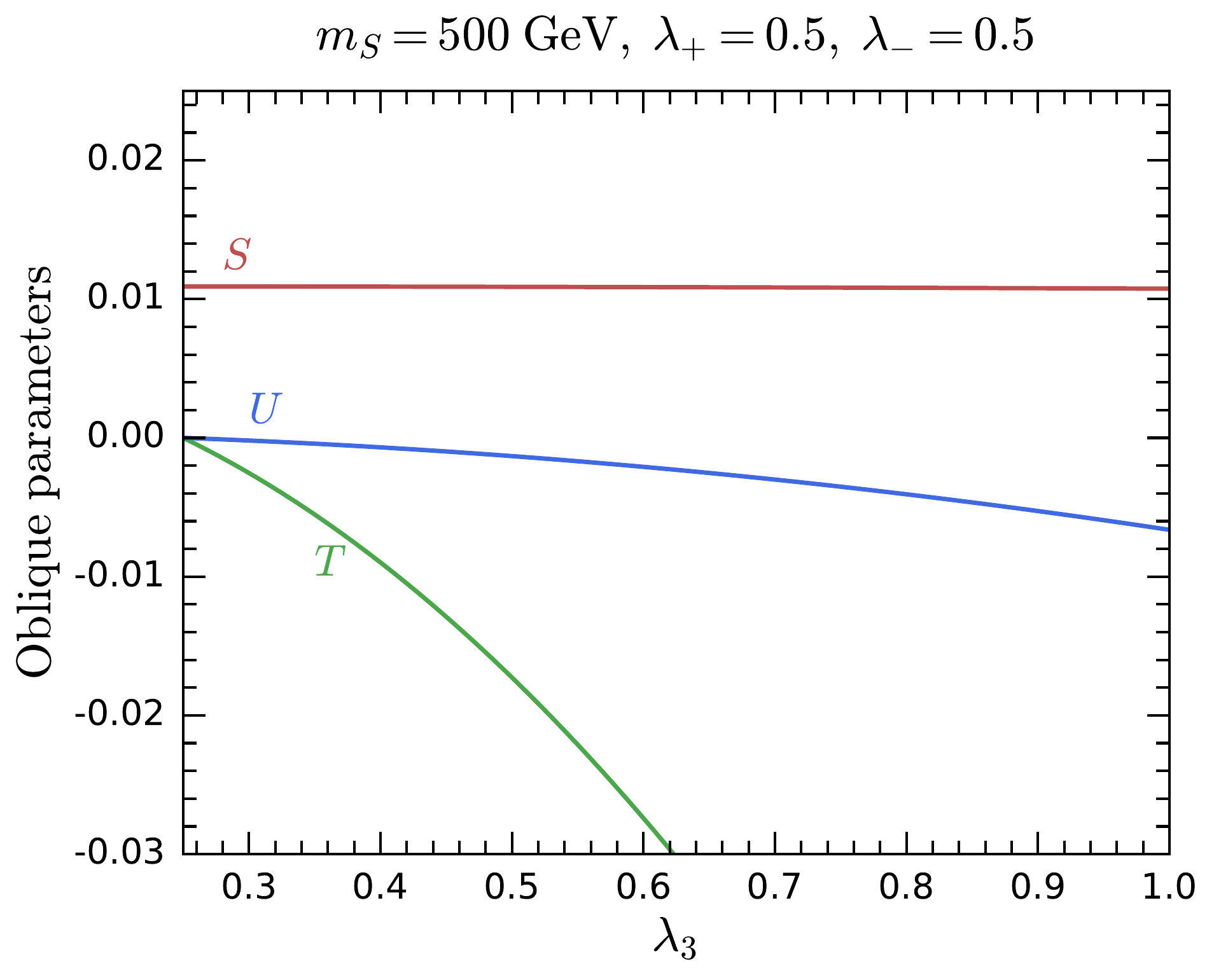}}
\caption{Electroweak oblique parameters $S$, $T$, and $U$ as functions of $\lambda_-$ for $\lambda_+ = \lambda_3 = 0.5$ (a) and of $\lambda_3$ for $\lambda_+ = \lambda_- = 0.5$ (b).
For both panels, $m_S = 500~\si{GeV}$.}
\label{fig:STU}
\end{figure}

In Fig.~\ref{fig:STU:lam-}, we plot the electroweak oblique parameters $S$, $T$, and $U$ as functions of $\lambda_{-}$ for $m_S = 500~\si{GeV}$ and $\lambda_+ = \lambda_3 = 0.5$.
In this case, $S$ increases as $\lambda_-$ increases, and $T$ and $U$ are not positive.
At $\lambda_- = 0$, $S$ vanishes, while $T$ and $U$ reach their minimums.
For $\lambda_- = \pm 1$, the custodial symmetry condition $\lambda_- = \pm 2\lambda_3$ is satisfied, leading to $T=U=0$.
In Fig.~\ref{fig:STU:lam3}, the oblique parameters are presented as functions of $\lambda_3$ for $m_S = 500~\si{GeV}$ and $\lambda_+ = \lambda_- = 0.5$.
In this case, $S$ hardly shows dependence on $\lambda_3$, while $T$ and $U$ decrease as $\lambda_3$ increases.
In both cases, $U$ is always quite small, and this is typical in electroweak multiplet DM models~\cite{Cai:2016sjz,Cai:2017wdu}.
Therefore, we neglect $U$ in the following analysis.

\begin{figure}[!t]
\centering
\subfigure[~$\lambda_+ = 1$, $\lambda_3 = 0.5$\label{fig:EWPT:lam-}]
{\includegraphics[width=0.49\textwidth]{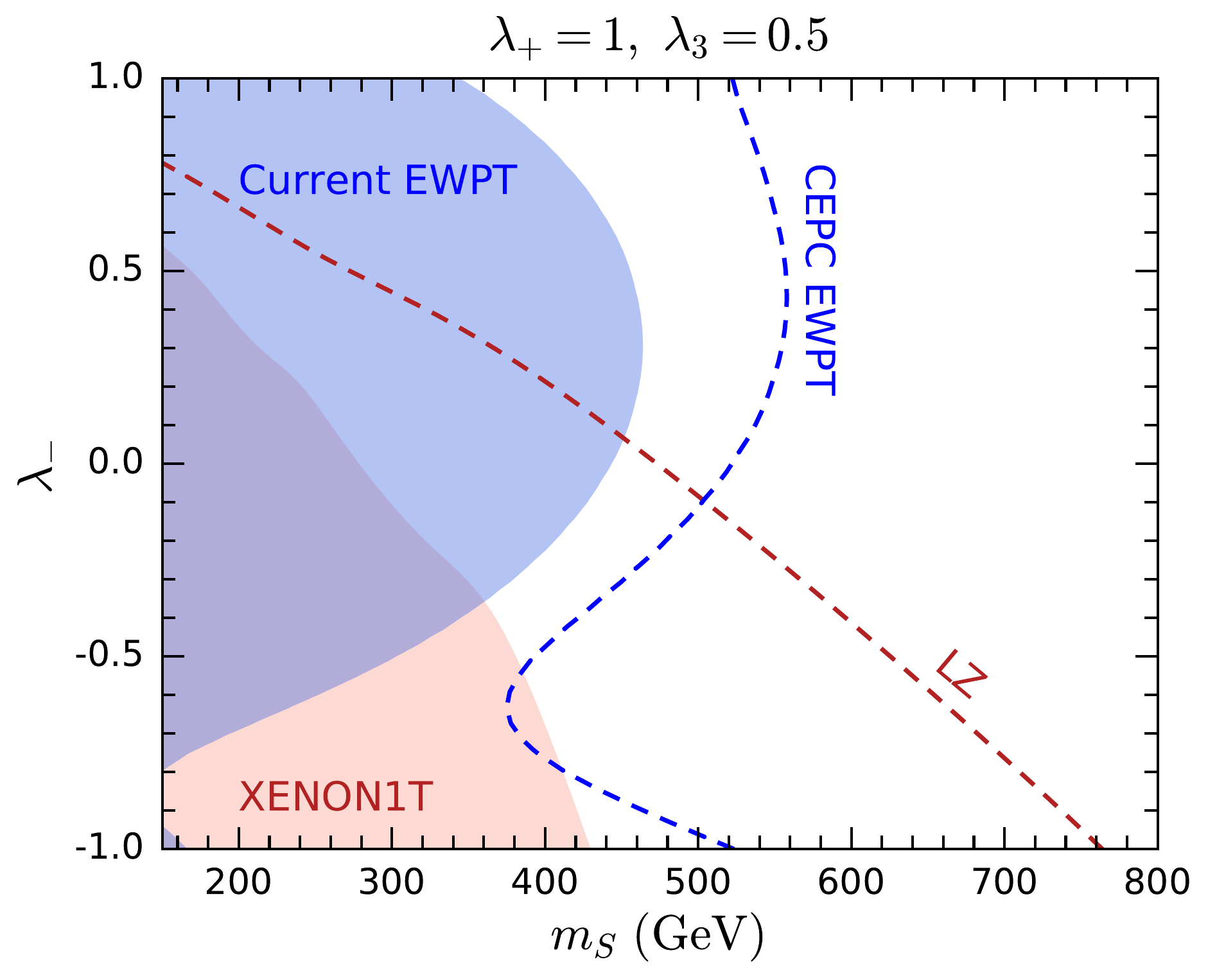}}
\subfigure[~$\lambda_+ = \lambda_- = 0.5$~\label{fig:EWPT:lam3}]
{\includegraphics[width=0.49\textwidth]{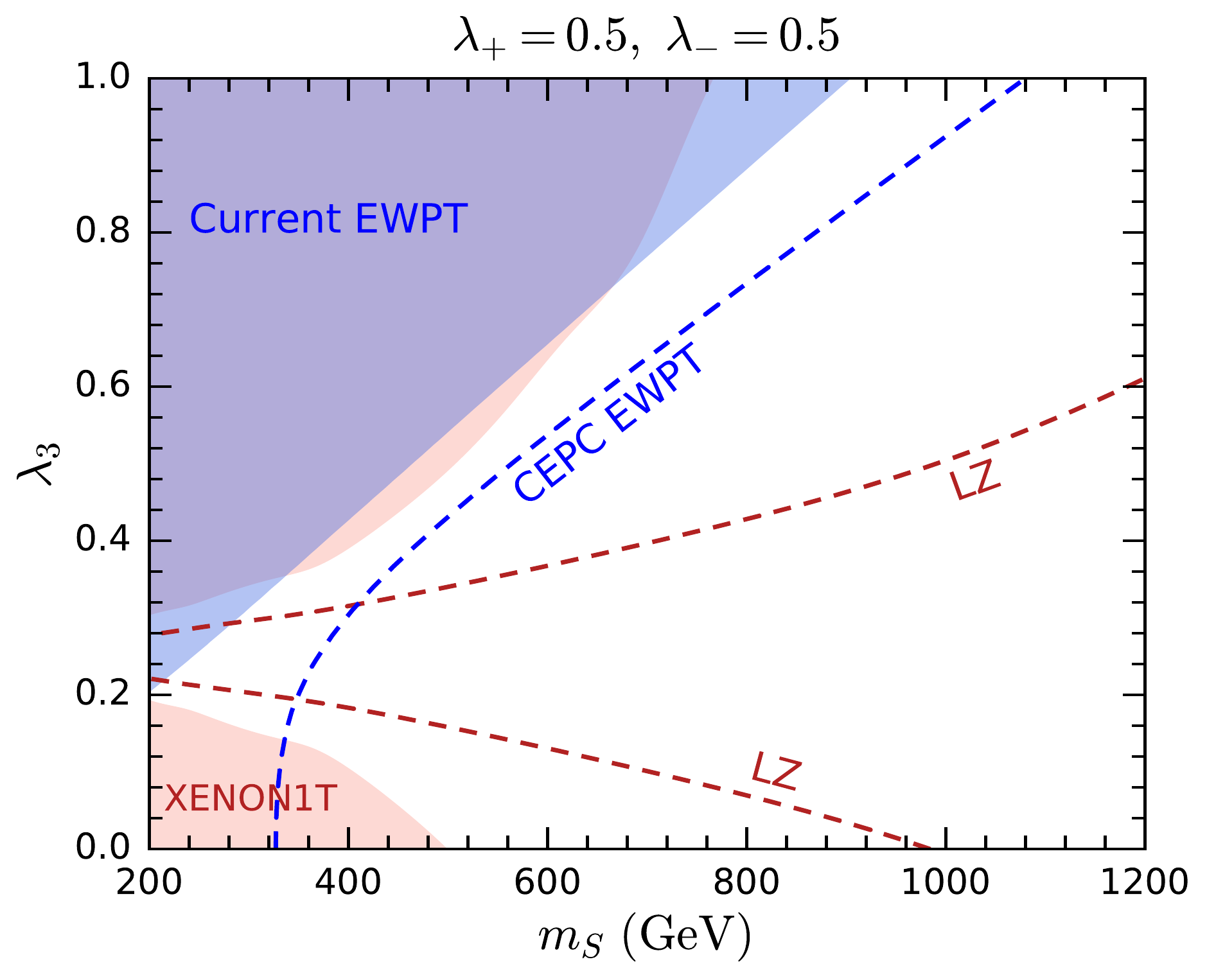}}
\caption{Current constraints and future sensitivities of EWPT and direct detection in the $m_S$-$\lambda_-$ plane for $\lambda_+ = 1$ and $\lambda_3 = 0.5$ (a) and in the $m_S$-$\lambda_3$ plane for $\lambda_+ = \lambda_- = 0.5$ (b).
The blue shade regions are excluded at 95\% C.L. by the global fit of current EWPT data from the Gfitter Group~\cite{Baak:2014ora}.
The blue dashed lines denote the expected 95\% C.L. exclusion limits of the future EWPT at the CEPC~\cite{CEPCStudyGroup:2018ghi}.
The red shade regions are excluded at 90\% C.L. by the XENON1T direct detection experiment~\cite{Aprile:2018dbl}.
The red dashed lines indicate the expected 90\% C.L. exclusion limits of the future LZ direct detection experiment~\cite{Mount:2017qzi}.}
\label{fig:EWPT}
\end{figure}

Assuming $U = 0$, the Gfitter Group performs a global fit of current EWPT data and gives~\cite{Baak:2014ora}
\begin{equation}
S = 0.06\pm 0.09,\quad
T = 0.10\pm 0.07,\quad
\rho_{ST} = 0.91,
\end{equation}
where $\rho_{ST}$ is the correlation coefficient between $S$ and $T$.
We use this result to constrain the inert sextuplet model, as shown in Fig.~\ref{fig:EWPT}.
We find that the current EWPT excludes a region with $m_S\lesssim 460~\si{GeV}$ in Fig.~\ref{fig:EWPT:lam-} for $\lambda_+ = 1$ and $\lambda_3 = 0.5$,
and also excludes a region up to $m_S\sim 900~\si{GeV}$ in Fig.~\ref{fig:EWPT:lam3} for  $\lambda_+ = \lambda_- = 0.5$.
Moreover, in Fig.~\ref{fig:LHC:soft:excluded} for the relation $\lambda_+ = \lambda_- = 2\lambda_3$, a region with $m_S \lesssim 140~\si{GeV}$ and $\lambda_3 \leq 0.5$ is excluded by the current EWPT.
Such a constraint is more stringent than the constraint from the 13~TeV monojet search, but looser than that from the 13~TeV soft-dilepton search.

Future $e^+e^-$ collider projects could greatly improve the measurement of the electroweak oblique parameters, providing a powerful indirect probe to the sextuplet scalar.
For instance, the CEPC EWPT could reach a precision of~\cite{CEPCStudyGroup:2018ghi}
\begin{equation}
\sigma_S = 0.01,\quad
\sigma_T = 0.01,\quad
\rho_{ST} = 0.62,
\end{equation}
where $\sigma_S$ and $\sigma_T$ are the $1\sigma$ uncertainties of $S$ and $T$, respectively.
Figure~\ref{fig:EWPT} shows the expected 95\% C.L. exclusion limits of the CEPC EWPT.
We find that CEPC could reach up to $m_S \sim 560~\si{GeV}$ in Fig.~\ref{fig:EWPT:lam-} and $m_S\sim 1.08~\si{TeV}$ in Fig.~\ref{fig:EWPT:lam3}.
In Fig.~\ref{fig:future:soft} for the relation $\lambda_+ = \lambda_- = 2\lambda_3$, the CEPC EWPT could probe a region with $m_S \lesssim 500~\si{GeV}$ and $\lambda_3 \leq 0.5$.

Now we discuss the constraint from DM direct detection experiments for comparison.
Unlike the collider analyses above, the interpretation of direct detection results depends on the DM density around the Earth, which is inferred from astrophysical observations.
For the inert sextuplet model, a related issue is whether the DM candidate $a^0$ is responsible for the whole cold dark matter.

Under the assumption of the conventional thermal production mechanism, the DM relic abundance calculation including the coannihilation and Sommerfeld enhancement effects in Ref.~\cite{Logan:2016ivc} showed that the observed relic abundance corresponds to $m_{a^0} \sim 6\text{--}17~\si{TeV}$ in the inert sextuplet model.
On the other hand, the future collider searches discussed above could can only probe mass scales up to $1\text{--}2~\si{TeV}$.
If we assume that dark matter solely originates from the thermal mechanism, the mass range that the future colliders can probe would lead to relic abundance lower than the observed value.
In this case, the DM particles from the inert sextuplet model could only constitute a fraction of dark matter, and the direct detection signals would be reduced by this fraction.

Nevertheless, if nonthermal production processes of the DM particles~\cite{Frieman:1989vt,Jeannerot:1999yn,Gherghetta:1999sw,Moroi:1999zb,Lin:2000qq,Fujii:2002kr,Kane:2015qea} also occur in the early Universe, it is still possible to obtain the observed relic abundance.
Here we assume that there are extra nonthermal production mechanisms to make $a^0$ particles comprise the whole cold DM.
Thus, we can interpret the direct detection constraint in the usual way.
In Fig.~\ref{fig:EWPT} we also demonstrate the regions excluded at 90\% C.L. by the XENON1T direct detection experiment~\cite{Aprile:2018dbl}, as well as the expected 90\% C.L. exclusion limits of the future LZ direct detection experiment~\cite{Mount:2017qzi}.
Note that the condition~\eqref{eq:flat_cond} corresponds to $\lambda_- = 1$ for $\lambda_+ = 1$ and $\lambda_3 = 0.5$ in Fig.~\ref{fig:EWPT:lam-} and to $\lambda_3 = 0.25$ for  $\lambda_+ = \lambda_- = 0.5$ in Fig.~\ref{fig:EWPT:lam3}.
Therefore, direct detection loses sensitivity in the regions around $\lambda_- \sim 1$ and $\lambda_3 \sim 0.25$, respectively.
Nonetheless, the CEPC EWPT provides a complementary approach to such regions.
We remark that if thermal production is the only mechanism that generates the DM particles in the early Universe, the direct detection constraint and sensitivity would be significantly weakened.

\section{Conclusions and discussions}
\label{sec:conslusion}

In this work, we have investigated a SM extension with an inert $\SUtwoL$ sextuplet scalar of hypercharge $1/2$.
After electroweak symmetry breaking, the quartic couplings between the sextuplet and the Higgs doublet split the sextuplet components.
Thus, the mass eigenstates in the dark sector include one triply charged scalar $S^{3\pm}$, two doubly charged scalars $S^{2\pm}_{1,2}$, two singly charged scalars $S^{\pm}_{1,2}$, and two neutral real scalars $\phi^0$ and $a^0$.
When the quartic couplings satisfy $\lambda_3\geq 0$ and $|\lambda_-|\leq 2\lambda_3$, $a^0$ is the lightest dark sector scalar, acting as a viable DM candidate.

The mass spectrum in this model is typically compressed, leading to 3-body decays of the dark sector scalars mediated by the $W$, $Z$, and Higgs bosons.
After pairs of these scalars are produced by electroweak processes at the LHC or a future 100~TeV $pp$ collider, the resulting decay products are typically soft.
This motivates us to consider the $\text{monojet} + \missET$ and $\text{soft-dilepton} + \text{jets} + \missET$ final states to search for the dark sector scalars at $pp$ colliders.
When the quartic couplings satisfy a particular relation \eqref{eq:flat_cond}, the DM-nucleon scattering is absent at tree level and DM direct detection experiments hardly probe the inert sextuplet model.
In this case, collider searches provide complementary approaches to this model.

Based on Monte Carlo simulation, we have derived the constraints from current LHC searches and have further evaluated the sensitivities at a 100~TeV $pp$ collider.
We have found that the 13~TeV monojet search excludes the parameter space up to $m_S \sim 85~\si{GeV}$, while the 13~TeV soft-dilepton search is more powerful, excluding the parameter space up to $m_S \sim 210~\si{GeV}$.
At a 100~TeV $pp$ collider, nonetheless, the monojet channel is much more sensitive than the soft-dilepton channel for probing the high mass region.
This is because the mass spectrum becomes more compressed at higher mass scales, leading to softer decay products.
With an integrated luminosity of $3~\si{ab^{-1}}$ at $\sqrt{s} = 100~\si{TeV}$, the monojet (soft-dilepton) search could explore the parameter space up to $m_S \sim 1.65~(0.84)~\si{TeV}$.

Since the sextuplet components contribute to the electroweak oblique parameters at one-loop level, EWPT provides an indirectly path to probe the model.
We have found that the current EWPT excludes a paramter region with $m_S \lesssim 900~\si{GeV}$ for $\lambda_+ = \lambda_- = 0.5$, while the future CEPC EWPT could probe the parameter space up to $m_S \sim 1.08~\si{TeV}$.

From the above analyses, we have found that the future collider searches can only explore mass scales up to $\sim 1.65~\si{TeV}$, where the thermal production of the DM particles leads to relic abundance lower than the observed value.
Therefore, if the $a^0$ particles in the inert sextuplet model exclusively originates from the thermal mechanism, they could only constitute a subdominant component of total dark matter in the parameter regions that the future colliders can reach.
Nevertheless, if some nonthermal production processes also occur in the early Universe, the $a^0$ particles with TeV masses might make up the whole cold DM.

\begin{acknowledgments}

This work is supported in part by the National Natural Science Foundation of China under Grants No.~11805288, No.~11875327, and No.~11905300, the China Postdoctoral Science Foundation
under Grant No.~2018M643282, the Natural Science Foundation of Guangdong Province
under Grant No.~2016A030313313,
the Fundamental Research Funds for the Central Universities,
and the Sun Yat-Sen University Science Foundation.

\end{acknowledgments}

\appendix

\section{Electroweak gauge couplings of the sextuplet scalar}
\label{app:gauge}

From Eqs.~\eqref{eq:L} and \eqref{eq:CD}, we derive the electroweak gauge couplings of the sextuplet scalar $S$.
The Lagrangian for the trilinear gauge couplings reads
\begin{eqnarray}
\mathcal{L}_{{\mathrm{gauge}}}^{{\mathrm{tri}}} &=& e{A_\mu }\big[3{({S^{3 + }})^\dag }i\overleftrightarrow {{\partial ^\mu }}{S^{3 + }} + 2{({S^{2 + }})^\dag }i\overleftrightarrow {{\partial ^\mu }}{S^{2 + }} + {({S^ + })^\dag }i\overleftrightarrow {{\partial ^\mu }}{S^ + } 
\nonumber\\
&&\quad~~~ - {({S^ - })^\dag }i\overleftrightarrow {{\partial ^\mu }}{S^ - } - 2{({S^{2 - }})^\dag }i\overleftrightarrow {{\partial ^\mu }}{S^{2 - }}\big]
\nonumber\\
&&  +\, \frac{g{Z_\mu }}{{2{c_{\mathrm{W}}}}}\big[(6c_{\mathrm{W}}^2 - 1){({S^{3 + }})^\dag }i\overleftrightarrow {{\partial ^\mu }}{S^{3 + }} + (4c_{\mathrm{W}}^2 - 1){({S^{2 + }})^\dag }i\overleftrightarrow {{\partial ^\mu }}{S^{2 + }} + (2c_{\mathrm{W}}^2 - 1){({S^ + })^\dag }i\overleftrightarrow {{\partial ^\mu }}{S^ + }
\nonumber\\
&& \quad \quad \quad ~  - {({S^0})^\dag }i\overleftrightarrow {{\partial ^\mu }}{S^0} - (2c_{\mathrm{W}}^2 + 1){({S^ - })^\dag }i\overleftrightarrow {{\partial ^\mu }}{S^ - } - (4c_{\mathrm{W}}^2 + 1){({S^{2 - }})^\dag }i\overleftrightarrow {{\partial ^\mu }}{S^{2 - }}\big]
\nonumber\\
&&  +\, g\bigg\{ \frac{{\sqrt {10} }}{2}W_\mu ^ + \big[{{({S^{3 + }})}^\dag }i\overleftrightarrow {{\partial ^\mu }}{S^{2 + }} + {{({S^ - })}^\dag }i\overleftrightarrow {{\partial ^\mu }}{S^{2 - }}\big] + 2W_\mu ^ + \big[{{({S^{2 + }})}^\dag }i\overleftrightarrow {{\partial ^\mu }}{S^ + } + {{({S^0})}^\dag }i\overleftrightarrow {{\partial ^\mu }}{S^ - }\big] 
\nonumber\\
&& \quad ~~ + \frac{{3\sqrt 2 }}{2}W_\mu ^ + {{({S^ + })}^\dag }i\overleftrightarrow {{\partial ^\mu }}{S^0} + \mathrm{H.c.} \bigg\},
\end{eqnarray}
where $\phi_1 \overleftrightarrow{\partial^\mu} \phi_2 \equiv \phi_1 \partial^\mu\phi_2 - (\partial^\mu\phi_1)\phi_2$.

The quartic gauge couplings are given by
\begin{eqnarray}
\mathcal{L}_{{\mathrm{gauge}}}^{{\mathrm{qua}}}  
&=& {e^2}{A_\mu }{A^\mu }\left( {9|{S^{3 + }}{|^2} + 4|{S^{2 + }}{|^2} + |{S^ + }{|^2} + |{S^ - }{|^2} + 4|{S^{2 - }}{|^2}} \right)
\nonumber\\
&&  + \frac{{{g^2}}}{{4c_{\mathrm{W}}^2}}{Z_\mu }{Z^\mu }\Big[ {{(6c_{\mathrm{W}}^2 - 1)}^2}|{S^{3 + }}{|^2} + {{(4c_{\mathrm{W}}^2 - 1)}^2}|{S^{2 + }}{|^2} + {{(2c_{\mathrm{W}}^2 - 1)}^2}|{S^ + }{|^2}
\nonumber\\
&& \quad \quad \quad \quad \quad~  + |{S^0}{|^2} + {{(2c_{\mathrm{W}}^2 + 1)}^2}|{S^ - }{|^2} + {{(4c_{\mathrm{W}}^2 + 1)}^2}|{S^{2 - }}{|^2} \Big]
\nonumber\\
&&  + \frac{{eg}}{{{c_{\mathrm{W}}}}}{A_\mu }{Z^\mu }\Big[ 3(6c_{\mathrm{W}}^2 - 1)|{S^{3 + }}{|^2} + 2(4c_{\mathrm{W}}^2 - 1)|{S^{2 + }}{|^2} + (2c_{\mathrm{W}}^2 - 1)|{S^ + }{|^2}
\nonumber\\
&&  \quad \quad \quad \quad \quad~ + (2c_{\mathrm{W}}^2 + 1)|{S^ - }{|^2} + 2(4c_{\mathrm{W}}^2 + 1)|{S^{2 - }}{|^2} \Big]
\nonumber\\
&&   +\, {g^2}W_\mu ^ + {W^{ - \mu }}\left[ {\frac{5}{2}(|{S^{3 + }}{|^2} + |{S^{2 - }}{|^2}) + \frac{{13}}{2}(|{S^{2 + }}{|^2} + |{S^ - }{|^2}) + \frac{{17}}{2}(|{S^ + }{|^2} + |{S^0}{|^2})} \right]
\nonumber\\
&&  +\, {g^2}\bigg( W_\mu ^ + {W^{ + \mu }}\Big\{ \sqrt {10} \big[{{({S^{3 + }})}^\dag }{S^ + } + {{({S^0})}^\dag }{S^{2 - }}\big] + 3\sqrt 2 \big[{{({S^{2 + }})}^\dag }{S^0} + {{({S^ + })}^\dag }{S^ - }\big] \Big\} + \mathrm{H.c.} \bigg)
\nonumber\\
&&   +\, g\Bigg( W_\mu ^ + \Bigg\{ \frac{{\sqrt {10} }}{2}\left[ {5e{A^\mu } + \frac{{g}}{{{c_{\mathrm{W}}}}}(5c_{\mathrm{W}}^2 - 1){Z_\mu }} \right]{{({S^{3 + }})}^\dag }{S^{2 + }} 
\nonumber\\
&& \quad \quad\quad \quad~~ + 2\left[ {3e{A_\mu } + \frac{{g}}{{{c_{\mathrm{W}}}}}(3c_{\mathrm{W}}^2 - 1){Z_\mu }} \right]{{({S^{2 + }})}^\dag }{S^ + }
\nonumber\\
&&  \quad \quad\quad \quad~~  + \frac{{3\sqrt 2 }}{2}\left[ {e{A_\mu } + \frac{{g}}{{{c_{\mathrm{W}}}}}(c_{\mathrm{W}}^2 - 1){Z_\mu }} \right]{({S^ + })^\dag }{S^0}
\nonumber\\
&& \quad \quad\quad \quad~~ - 2\left[ {e{A_\mu } + \frac{{g}}{{{c_{\mathrm{W}}}}}(c_{\mathrm{W}}^2 + 1){Z_\mu }} \right]{({S^0})^\dag }{S^ - }
\nonumber\\
&&  \quad \quad\quad \quad~~  - \frac{{\sqrt {10} }}{2}\left[ {3e{A_\mu } + \frac{{g}}{{{c_{\mathrm{W}}}}}(3c_{\mathrm{W}}^2 + 1){Z_\mu }} \right]{{({S^ - })}^\dag }{S^{2 - }} \Bigg\} + \mathrm{H.c.} \Bigg) .
\end{eqnarray}

\section{Vacuum polarizations of electroweak gauge bosons}
\label{app:Pi}

In this appendix, we explicitly list the one-loop contributions to gauge boson vacuum polarizations $\Pi_{PQ}(p^2)$ by the dark sector scalar bosons in the inert sextuplet model.
We express the results with the Passiano–Veltman scalar functions $A_0$ and $B_{00}$~\cite{Passarino:1978jh}, whose definitions are consistent with Ref.~\cite{Denner:1991kt}.
In our calculation, the numerical values of these functions are provided by \texttt{LoopTools}~\cite{Hahn:1998yk}.

The photon vacuum polarization contributed by the dark sector scalars is
\begin{eqnarray}
\Pi_{AA}(p^2) &=& \frac{e^{2}}{16\pi^{2}}\big\{
18[2B_{00}(p^2,m_{S^{3\pm} }^{2},m_{S^{3\pm}}^{2})
- A_{0}(m_{S^{3\pm} }^{2})]
\nonumber\\
&& ~
+8[2B_{00}(p^2,m_{S^{2\pm}_{1}}^{2},m_{S^{2\pm}_{1}}^{2})
-A_{0}(m_{S^{2\pm}_{1}}^{2})
+ 2B_{00}(p^2,m_{S^{2\pm}_{2}}^{2},m_{S^{2\pm}_{2}}^{2})
-A_{0}(m_{S^{2\pm}_{2}}^{2})]
\nonumber\\
&& ~ + 2[2B_{00}(p^2,m_{S^\pm_{1}}^{2},m_{S^\pm_{1}}^{2})
-A_{0}(m_{S^\pm_{1}}^{2})  +2B_{00}(p^2,m_{S^\pm_{2}}^{2},m_{S^\pm_{2}}^{2})
-A_{0}(m_{S^\pm_{2}}^{2})]\big\}.\qquad
\end{eqnarray}
The photon-$Z$ vacuum polarization reads
\begin{eqnarray}
\Pi_{AZ}(p^2) &=& \frac{e^{2}}{16\pi^{2}s_\mathrm{W}c_\mathrm{W}}\big\{
3(5-6s_\mathrm{W}^{2})[2 B_{00}(p^2,m_{S^{3\pm} }^{2},m_{S^{3\pm} }^{2})
 - A_{0}(m_{S^{3\pm}}^{2})]
\nonumber\\
&& \qquad\qquad\quad~ +2(3+2s_{2+}^{2}-4s_\mathrm{W}^{2})[2B_{00}(p^2,m_{S^{2\pm}_{1}}^{2},m_{S^{2\pm}_{1}}^{2})
-A_{0}(m_{S^{2\pm}_{1}}^{2})]
\nonumber\\
&& \qquad\qquad\quad~ +2(5-2s_{2+}^{2}-4s_\mathrm{W}^{2})  [2B_{00}(p^2,m_{S^{2\pm}_{2}}^{2},m_{S^{2\pm}_{2}}^{2})
- A_{0}(m_{S^{2\pm}_{2}}^{2})]
\nonumber\\
&& \qquad\qquad\quad~ +(1+2s_{+}^{2}-2s_\mathrm{W}^{2})[2B_{00}(p^2,m_{S^\pm_{1}}^{2},m_{S^\pm_{1}}^{2}) 
-A_{0}(m_{S^\pm_{1}}^{2})]
\nonumber\\
&& \qquad\qquad\quad~ +(3-2s_{+}^{2}-2s_\mathrm{W}^{2})  [2B_{00}(p^2,m_{S^\pm_{2}}^{2},m_{S^\pm_{2}}^{2})
 - A_{0}(m_{S^\pm_{2}}^{2})] \big\},
\end{eqnarray}
where $s_{2+}\equiv \sin\theta_{2+}$ and $s_{+}\equiv \sin\theta_{+}$.
The $Z$ boson vacuum polarization is given by
\begin{eqnarray}
\Pi_{ZZ}(p^2) &=& \frac{e^{2}}{32\pi^{2}s_\mathrm{W}^{2}c_\mathrm{W}^{2}}\big\{
2(5-6s_\mathrm{W}^{2})^{2}                     B_{00}(p^2,m_{S^{3\pm}}^{2},m_{S^{3\pm}}^{2})
+                                      2B_{00}(p^2,m_{\phi^{0}}^{2},m_{a^{0}}^{2})
\nonumber\\
&& \quad~+2(3+2s_{2+}^{2}-4s_\mathrm{W}^{2})^{2}B_{00}(p^2,m_{S^{2\pm}_{1}}^{2},m_{S^{2\pm}_{1}}^{2}) 
+16s_{2+}^{2}c_{2+}^{2}B_{00}(p^2,m_{S^{2\pm}_{1}}^{2},m_{S^{2\pm}_{2}}^{2})
\nonumber\\
&&
\quad~+2(5-2s_{2+}^{2}-4s_\mathrm{W}^{2})^{2}B_{00}(p^2,m_{S^{2\pm}_{2}}^{2},m_{S^{2\pm}_{2}}^{2})
\nonumber\\
&&\quad~
+2(1+2s_{+}^{2}-2s_\mathrm{W}^{2})^{2}B_{00}(p^2,m_{S^\pm_{1}}^{2},m_{S^\pm_{1}}^{2}) 
+16s_{+}^{2}c_{+}^{2}B_{00}(p^2,m_{S^\pm_{2}}^{2},m_{S^\pm_{1}}^{2})
\nonumber\\
&&\quad~
+2(3-2s_{+}^{2}-2s_\mathrm{W}^{2})^{2}B_{00}(p^2,m_{S^\pm_{2}}^{2},m_{S^\pm_{2}}^{2}) 
\nonumber\\
&&\quad~ -(5-6s_\mathrm{W}^2)^2 A_{0}(m_{S^{3\pm}}^{2})
-A_{0}(m_{\phi^{0}}^{2})/2-A_{0}(m_{a^{0}}^{2})/2
\nonumber\\
&&\quad~ - [(3-4s_\mathrm{W}^2)^2 c_{2+}^2 + (5-4s_\mathrm{W}^2)^2 s_{2+}^2]A_{0}(m_{S^{2\pm}_{1}}^{2}) 
\nonumber\\
&&\quad~ - [(3-4s_\mathrm{W}^2)^2 s_{2+}^2 + (5-4s_\mathrm{W}^2)^2 c_{2+}^2] A_{0}(m_{S^{2\pm}_{2}}^{2})  
\nonumber\\
&&\quad~ - [(1-2s_\mathrm{W}^2)^2 c_{+}^2 + (3-2s_\mathrm{W}^2)^2 s_{+}^2] A_{0}(m_{S^\pm_{1}}^{2}) 
\nonumber\\
&&\quad~ - [(1-2s_\mathrm{W}^2)^2 s_{+}^2 + (3-2s_\mathrm{W}^2)^2 c_{+}^2] A_{0}(m_{S^\pm_{2}}^{2})
\big\},
\end{eqnarray}
where $c_{2+}\equiv \cos\theta_{2+}$ and $c_{+}\equiv \cos\theta_{+}$.
The $W$ boson vacuum polarization is
\begin{eqnarray}
\Pi_{WW}(p^2) &=& \frac{e^{2}}{16\pi^{2}s_\mathrm{W}^{2}}\big[
10c_{2+}^{2}                                              B_{00}(p^2,m_{S^{3\pm}}^{2} ,m_{S^{2\pm}_{1}}^{2})
+10s_{2+}^{2}                                                     B_{00}(p^2,m_{S^{3\pm}}^{2} ,m_{S^{2\pm}_{2}}^{2})
\nonumber\\
&& \quad~+4(2c_{+}c_{2+}-\sqrt{5/2}\,s_{+}s_{2+})^{2}B_{00}(p^2,m_{S^{2\pm}_{1}}^{2},m_{S^\pm_{1}}^{2}) 
\nonumber\\
&& \quad~+4(2s_{+}c_{2+}+\sqrt{5/2}\,c_{+}s_{2+})^{2}B_{00}(p^2,m_{S^{2\pm}_{1}}^{2},m_{S^\pm_{2}}^{2})  
\nonumber\\
&& \quad~+4(2c_{+}s_{2+}+\sqrt{5/2}\,s_{+}c_{2+})^{2}B_{00}(p^2,m_{S^{2\pm}_{2}}^{2},m_{S^\pm_{1}}^{2}) 
\nonumber\\
&&
\quad~+4(2s_{+}s_{2+}-\sqrt{5/2}\,c_{+}c_{2+})^{2}B_{00}(p^2,m_{S^{2\pm}_{2}}^{2},m_{S^\pm_{2}}^{2})  
\nonumber\\
&& \quad~+4(3c_{+}/2-\sqrt{2}\,s_{+})^{2}                          B_{00}(p^2,m_{S^\pm_{1}}^{2},m_{\phi^{0}}^{2})
\nonumber\\
&& \quad~+4(3c_{+}/2+\sqrt{2}\,s_{+})^{2}                            B_{00}(p^2,m_{S^\pm_{1}}^{2},m_{a^{0}}^{2})  
\nonumber\\
&& \quad~+4(3s_{+}/2+\sqrt{2}\,c_{+})^{2}                            B_{00}(p^2,m_{S^\pm_{2}}^{2},m_{\phi^{0}}^{2}) 
\nonumber\\
&& \quad~+4(3s_{+}/2-\sqrt{2}\,c_{+})^{2}                            B_{00}(p^2,m_{S^\pm_{2}}^{2},m_{a^{0}}^{2})
\nonumber\\
&& \quad~-5A(m_{S^{3\pm}}^{2})/2
-({13}/{2}-4s_{2+}^{2})A_{0}(m_{S^{2\pm}_{1}}^{2})-(5/2+4s_{2+}^{2})A_{0}(m_{S^{2\pm}_{2}}^{2}) 
\nonumber\\
&& \quad~-({17}/{2}-2s_{+}^{2})A_{0}(m_{S^\pm_{1}}^{2})-({13}/{2}+2s_{+}^{2})A_{0}(m_{S^{\pm}_{2}}^{2})
\nonumber\\
&& \quad~-17A_{0}(m_{\phi^{0}}^{2})/4-17A_{0}(m_{a^{0}}^{2})/4\big].
\end{eqnarray}

\bibliographystyle{utphys}
\bibliography{ref}
\end{document}